\documentclass[authoryear,preprint,review,12pt]{elsarticle}

\usepackage{fullpage}
\usepackage{graphicx} 
\usepackage[export]{adjustbox} 

\usepackage{booktabs} 
\usepackage{graphicx} 
\usepackage[table,xcdraw]{xcolor} 

\usepackage{subcaption}

\usepackage[utf8]{inputenc}
\usepackage[english]{babel}
\usepackage{amsmath}
\usepackage{amsfonts}
\usepackage{amssymb}
\usepackage{xcolor}
\usepackage{epstopdf}
\usepackage{ragged2e}
\justifying
\usepackage{multicol}
\usepackage{lmodern}

\usepackage{wrapfig}
\usepackage{graphicx}
\usepackage{float}
\graphicspath{{./Figures}{Figures/LES}{Figures/LES/sampleT_27}{Figures/LES/sampleT_27}{Figures/allResults/figsCrop/Ex-1-1-1}{Figures/allResults/figsCrop/Ex-1-1-2}}

\usepackage{lmodern}
\usepackage{epstopdf}

\usepackage{hyperref}
\hypersetup{
    colorlinks=true,
    linkcolor=blue,
    filecolor=magenta,      
    urlcolor=cyan,
    pdftitle={Overleaf Example},
    pdfpagemode=FullScreen,
    }

\usepackage{fancyvrb}
\begin{document}

\begin{frontmatter}



\title{Uncertainty Quantification and Flow Dynamics in Rotating Detonation Engines}

\author[1]{Vinay Kumar}
\author[1]{Zheming Gou\corref{cor1}}
\author[1,2]{Roger Ghanem}

\cortext[cor1]{Corresponding author. Email: zgou@usc.edu}

\address[1]{Department of Aerospace \& Mechanical Engineering, University of Southern California, Los Angeles, CA 90089, USA}
\address[2]{Department of Civil \& Environmental Engineering, University of Southern California, Los Angeles, CA 90018, USA}

\begin{abstract}
Rotating detonation engines (RDEs) are a critical technology for advancing combustion engines, particularly in applications requiring high efficiency and performance. Understanding the supersonic detonation structure and how various parameters influence these phenomena is essential for optimizing RDE design. In this study, we perform detailed simulations of detonations in an RDE and analyze how the flow patterns are affected by key parameters associated with the droplet arrangement within the engine. To further explore the system's sensitivity, we apply polynomial chaos expansion to investigate the propagation of uncertainties from input parameters to quantities of interest (QOIs). Additionally, we develop a framework to accurately characterize the joint distributions of QOIs with a limited number of simulations. Our findings indicate that the strategic release of droplets may be crucial for sustaining continuous detonation waves in the engine, and accurate representations of the solution (e..g via high-order chaos expansions) are essential to accurately capture the dependence between QOIs and input uncertainties. These insights provide a quantitative foundation for further optimization of RDE designs.
\end{abstract}



\begin{keyword}
Rotation detonation engine \sep Hypersonic combustion \sep Uncertainty quantification \sep Polynomial Chaos expansion



\end{keyword}

\end{frontmatter}

\section{Introduction:}

\subsection{Importance and challenges for complete RDE setup}

Combustion engines, where heat is produced by a chemical reaction between a fuel and an oxidizer after deflagration or subsonic combustion, have been an ubiquitous source of energy generation and propulsion for several decades.
Detonative combustion, a supersonic combustion process initially proposed a few decades ago \cite{nicholls1957intermittent}, has recently received renewed and growing attention. Detonation combustion involves the rapid conversion of chemical energy into mechanical energy by the formation and propagation of shock waves using a fuel-air mixture. Rapid and efficient energy release results in improved engine performance and efficiency compared to conventional combustion engines.
The practical implementation of detonative combustion is hampered by a number of challenges, including confinement conditions such as temperature and pressure. 
The optimized engine design and operating conditions are promoted by such factors as engine geometry, fuel injection strategies, and turbulence enhancement techniques.
Among the innovative concepts emerging in recent years, the rotating detonation engine (RDE) stands out as a promising candidate that offers higher efficiency, increased power output, and reduced fuel consumption compared to conventional propulsion methods~\citep{raman2023nonidealities, rankin2017overview, kailasanath2011rotating}.
%
Moreover, by harnessing the rapid and efficient energy release associated with detonation, RDE has the potential to significantly improve propulsion systems' performance in a number of applications, including aerospace, automotive, and power generation.

\subsection{Main component: sustainability of moving detonation wave}
A key and the distinguishing feature of RDE is the circumferentially moving detonation shock waves. To satisfy operational conditions, these wavelets are required to be continuously maintained, proffering the required fuel proportions in the combustor. Such fuel availability depends upon several factors chief among which are the fuel-injection characteristics. 
\cite{ma2018experimental}
carried out an experimental investigation on hydrogen-air rotating detonation engine performance subjected to ignition,
quenching, re-initiation, and stabilization.
An array of fuel injection inlets, into the combustor, was used. The experimental investigation showed the existence of transitions between single and multiple detonation waves.
More specifically, it was noticed that robust detonation waves weaken their intensity gradually and eventually vanish. On the other hand, it was also observed that a weaker wave appears due to high pressure and temperature spots created by shock wave reflections. 
The hotspot for such behavior was noticed near the injection sections of the RDE combustor. 
Furthermore, \cite{wu2015stability} observed unstable trajectories of the transverse wave as well as switching between single spin and its associated higher modes.
\cite{gordon1958limit} commented on detonation limits coupled with reaction kinetic and heat loss effects.  

\subsection{Inlet geometries in literature and present study}
\begin{figure}
    \centering
    \includegraphics[width=0.6\linewidth]{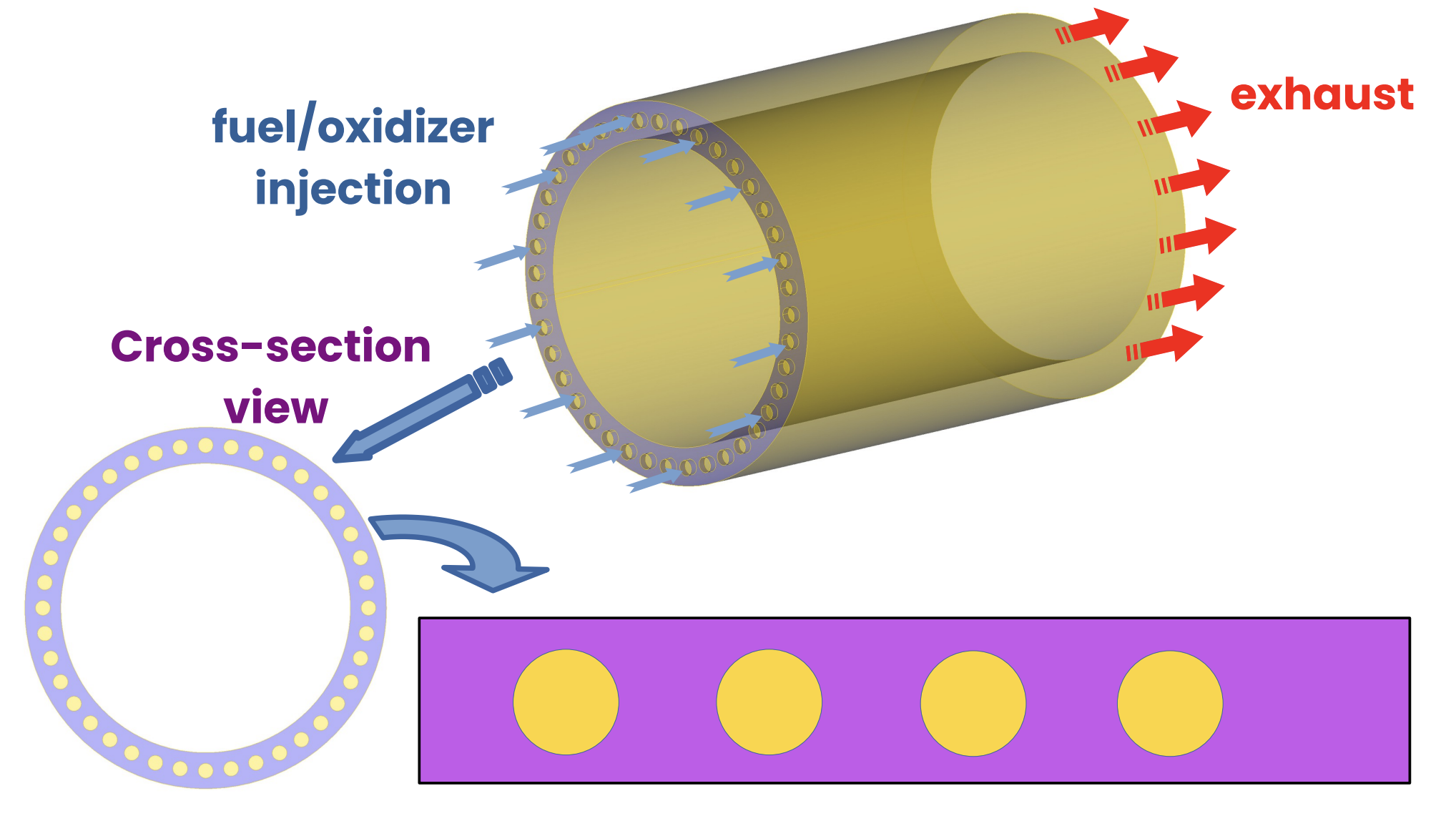}
    \caption{The schematic diagram of inlets in rotating detonation combuster.}
    \label{fig:rotation engine}
\end{figure}

The geometry of RDE~\citep{kindracki2020influence} consists of a hollow cylindrical shape (space between two concentric cylinders) for detonative combustion to take place. Here, fuel is injected in the axial direction from one annular section, and the hot products released from combustive reactions are assumed to be propelled through another annular section. The spinning detonation waves are assumed to travel in the azimuthal direction. 
In 2D flow simulations, as illustrated in Figure~\ref{fig:rotation engine}, the axial direction is considered the vertical axis, and the azimuthal direction of the annular section is represented as the horizontal axis. 
More importantly, the radial direction is neglected because the effects are not assumed to be worth considering due to the small length in comparison to axial and azimuthal directions. 
However, the thin radial section induces strong detonative reflections of the wave, which can play a significant role in spinning detonation wavelet dynamics. 

Since the fuel injection is in the axial direction, inlets are required on the annular section. Modeling these can proceed in one of two ways: a) every boundary point can be considered as a micro-nozzle and 2) a finite number of spatially discrete slot injectors is introduced. The second approach is closer to an operational setup ~\citep{palaniswamy2018comparison,sun2017effects,schwer2011effect,betelin20203d}.
In a number of such studies \cite{yan2021effects, bluemner2020effect}, it was concluded that geometrical variations in spatially discrete injection slots significantly affect the characteristics of rotating detonation wave propagation. More specifically, these studies drew an inference that irregularities due to such spatial discrete placement of fuel inlets dominate the detonation front instability. 

\subsection{Highlights}
The limitations of deflagration combustion ~\citep{humphrey1909internal,roy1946propulsion, nicholls1957intermittent, varsi1976blast, back1975application} have been increasingly in focus, in tandem with demands on efficiency, economy,and functionality. While RDE provides an appealing resolution to these operational challenges, it presents an important  technical challenge. Specifically, maintaining the detonation wave for a sufficiently long time remains elusive as the detonative shock wave depends on several factors for its stability. Foremost among these factors are the characteristics of fuel injection into the combustor, with the key characteristics being  randomness, inlet geometry and spatial pattern of axial injection~\cite {lu2014rotating}. These parameters are translated into randomness in fuel availability to the combustion process.
Informed by this insight, the present study explores in a systematic fashion the effect of geometry and size of inlets as well as fuel composition on the combustion process.
Moreover, when fuel comes into high pressure and temperature conditions, exothermic reactions occur. Consequently, the released heat increases temperature and, subsequently, raises the pressure of the enclosed domain. Higher pressure results in higher density flow and higher velocity. Clearly, the gradients of concentration of fuel components, of temperature, pressure, density, and velocity are highly correlated. Based on such observations, we choose gradients of pressure as the main quantities of interest (QOIs). To assess the velocity characteristics, we visualize and study Q-criteria and vorticity. 
%
To showcase the effects of uncertainties in the inlet geometries, special alignment, and fuel composition, a mathematical model for the compressible Navier-Stokes equations coupled with heat and concentration is used. Statistical analysis is carried out by constructing polynomial chaos expansion (PCE) of the QOIs with respect to the above-mentioned random input.

\section{Governing Equations:}
The detonative combustion in the present study is investigated using mathematical modeling of a compressible reactive system. 
When we denote the conservative variable as $\mathbf{U}$, fluxes as $\mathbf{F}$, and source terms as $\mathbf{S}$, the equations governing such flow dynamics are as follows
\begin{equation}\label{eqn:main}
\partial_{t} \mathbf{U} + \nabla \cdot \mathbf{F} = \mathbf{S}    
\end{equation}
where 
\begin{equation}\label{eqn:U_F_S}
\begin{split}
\mathbf{U} = 
	\left( \begin{array}{c}
	\rho \\
	\rho Y_{k} \\
	\rho \mathbf{u} \\
	\rho E 
	\end{array} \right) , 
\quad
\mathbf{F} = 
	\left( \begin{array}{c}
	\rho \mathbf{u} \\
	\rho Y_{k} \mathbf{u}\\
	\rho \mathbf{u} \otimes \mathbf{u} + p \mathbf{I} \\
	(\rho E + p) \mathbf{u}
	\end{array} \right) ,
\quad 	
\mathbf{S} = 
	\left( \begin{array}{c}
	0 \\
	\dot{R}_{k} + \nabla \cdot \mathbf{J}_{k} \\
	\dot{M} + \dot{M}_{\nu} \\
	\dot{E} + \dot{E}_{\nu} 
	\end{array} \right) 
\end{split}\ .
\end{equation}
Here, $\rho$, $\mathbf{u}$, $E$, and $p$ represent density, velocity, total energy (thermal and kinetic), and pressure, respectively. $Y_{k}$ and $\mathbf{J}_{k}$ are $k$-th specie's mass fraction and diffusion fluxes. $\dot{R}_{k}$, $\dot{M}$, and $\dot{E}$ are the source terms for species production from reactions, momentum, and energy sources. $\dot{M}_{\nu}$ and $\dot{E}_{\nu}$ are viscous terms.
For the inviscid case $\dot{M}_{\nu} = \dot{E}_{\nu} = 0$. 
In the case of viscous flow, viscosity is included as follows,
\begin{equation*}
\dot{M}_{\nu} = \nabla \cdot \left( \rho \nu \left[ \nabla \mathbf{u} + \left( \nabla \mathbf{u} \right)^{T} - \frac{2}{3} \left( \nabla \cdot \mathbf{u} \right) \mathbf{I} \right] \right)
\end{equation*}	

\begin{equation*}
\dot{E}_{\nu} = \nabla \cdot \left( \rho \nu \left[ \nabla \mathbf{u} + \left( \nabla \mathbf{u} \right)^{T} - \frac{2}{3} \left( \nabla \cdot \mathbf{u} \right) \mathbf{I}  \right] \cdot \mathbf{u} \right) + \nabla \cdot \left( \alpha_{e} \nabla e \right)
\end{equation*}
$\nu$ and $\alpha_{T}$ denote viscosity and thermal diffusivity, respectively, while $\mathbf{I}$ denotes the identity matrix.





The total energy in Eq.~\eqref{eqn:U_F_S} can be written as $E = h_s - p\rho^{-1} + \frac{1}{2} \Vert \overrightarrow{u_k} \Vert ^{2}$ 
where the sensible enthalpy ($h_s$) is expressed in terms of temperature as $h_s = \int^{T}_{T_{0}} c_{p} dT$. Here, it is assumed that the reactants are premixed to behave as an ideal gas. Moreover, pressure $p$ is computed using the equation $p = \rho T \sum^{N}_{k=1}Y_{k}W^{-1}_{k}$, where $W_k$ denotes the molar mass of $k$-th species. The expression $\alpha_{T} \nabla e$ in energy equation Eq.~\eqref{eqn:U_F_S} represents the diffusive heat flux.


The heat generated during combustion is denoted by $\omega_T$, and $\dot{R}_k$ is $k$-th species's reaction rate. These are evaluated as
\begin{equation*}
\dot{\omega}_T = \sum^{N}_{k=1} \Delta h^{0}_{f,k} \dot{R}_k
\end{equation*}
\begin{equation*}
\dot{R}_{k} = W_{k}\sum^{N}_{k=1} \dot{q}_{kj}
\end{equation*}
where $\Delta h^{0}_{f,k}$ and $h_{s,k}$ are formation enthalpy and sensible enthalpy of $k$-th species, respectively. The mathematical relation of sensible enthalpy with temperature can be written as:
\begin{equation*}
h_{s,k} = \int^{T}_{T_{0}} c_{p,k} dT
\end{equation*}
In the combustion modeling of the present work, a detailed hydrogen-oxygen chemical kinetics mechanism is considered. The rates of fuel consumption and product formation are accounted for as source terms in the concentration equations. 
The stoichiometric equation for the chemical mechanism consisting of $m$ reactions and $n$ chemical species is expressed as:
\begin{equation*}
\sum_{k=1}^{n} v'_{kj} \, \mathrm{M}_k \;\leftrightarrow\; \sum_{k=1}^{n} v''_{kj} \, \mathrm{M}_k, \quad j = 1, 2, \dots, m
\end{equation*}
where $v'_{kj}$ and $v''_{kj}$ are the stoichiometric coefficients of species $k$ as a reactant and product in the $j$-th reaction, respectively, and $\mathrm{M}_k$ denotes the $k$-th chemical species. The net production rate of species $k$ due to the $j$-th reaction, denoted by $\dot{q}_{kj}$, is given by:
\begin{equation} \label{eq:q}
\dot{q}_{kj} = (v''_{kj} - v'_{kj}) \left[ k_{fj} \prod_{i=1}^{n} C_i^{v'_{ij}} - k_{bj} \prod_{i=1}^{n} C_i^{v''_{ij}} \right],
\end{equation}
where $C_i$ is the molar concentration of species $i$, and $k_{fj}$ and $k_{bj}$ are the forward and backward rate constants (units depend on reaction order). 
The total net production rate of species $k$ from all reactions is:
\begin{equation*}
\dot{q}_k = \sum_{j=1}^{m} \dot{q}_{kj}
\end{equation*}

\section{Polynomial chaos expansion}

Polynomial chaos expansion has been proposed as a powerful tool for uncertainty quantification and stochastic analysis~\citep{soize2004physical}. It is widely used in various engineering applications, including material characterization~\citep{spanos1989stochastic}, fluid flows~\citep{knio2001stochastic,le2002stochastic}, reaction chemistry~\citep{reagana2003uncertainty}, fluid-structure coupling problems~\citep{ghiocel2002stochastic} etc. This section describes the basic theory of PCE and explains the procedures to construct PCE.

\subsection{PCE basics}

Consider a random variable $\mathbf{X}\in\mathbb{R}^d$ defined on a probability triple $(\Omega, \mathcal{F}, \mathbb{P})$. Let $\mathbf X$ have the distribution function $F_{\mathbf{X}}(\mathbf{x})$ and associated density function $p_{\mathbf X}(\mathbf x)$. Consider the Hilbert space $\mathbb{H}$ of square-integrable functions $f:\mathbb{R}^d \rightarrow \mathbb{R}^m$ with respect to $F_{\mathbf X}$. Introduce the following inner product on $\mathbb{H}$,
\begin{equation*}
    \langle f, g\rangle_{\mathbb{H}} = \int  \langle f(\mathbf{x}),g(\mathbf{x})\rangle_{\mathbb{R}^m}dF_{\mathbf{X}}(\mathbf{x}) \ , \quad \forall f,g\in\mathbb{H}
\end{equation*}
where $\langle .,.\rangle_{\mathbb{R}^m}$ is the usual inner product in ${\mathbb{R}^m}$.
Let $\{\psi_{\boldsymbol\alpha}(\mathbf{x}), \boldsymbol\alpha\in\mathbb{R}^m\}$ be a complete orthonormal basis in $L^2(\mathbb{R}^d,\mathbb{R},F_{\mathbf X}(\mathbf x))$ obtained by taking the tensor product of a basis set in $L^2(\mathbb{R},\mathbb{R},F_{\mathbf X}(\mathbf x))$. This would be the case for instance when using multidimensional polynomials obtained as products of their one-dimensional counterparts. Next, we consider a physics model $\mathcal{M}$ that takes as input the $d$-dimensional parameters $\mathbf X$ and outputs an $m$-dimensional variable $\mathbf Y$.  Assuming the model $\mathcal{M}$ to be square-integrable, it is in $\mathbb{H}$. It then has the componentwise representation using the basis set $\{\psi_{\boldsymbol\alpha}\}$ \citep{soize2004physical}
\begin{equation}
    \label{eq:PCE}
    \mathbf Y = \mathcal{M}(\mathbf{X}) = \sum_{\boldsymbol\alpha\in\mathbb{N}^d}\mathbf Y_{\boldsymbol\alpha} \psi_{\boldsymbol\alpha}(\mathbf{X}) \ , \qquad \mathbf Y_{\boldsymbol\alpha}\in\mathbb{R}^m.
\end{equation}
In the case where $\mathbf{X}$ is a multivariate Gaussian vector, the functions $\psi_{\boldsymbol\alpha}$ can be selected as the normalized multivariate Hermite polynomials, and the spectral expansion \eqref{eq:PCE} is referred to as the polynomial chaos expansion. It is worth mentioning that mean-square convergence of \eqref{eq:PCE} is guaranteed under the above assumptions. 

The infinite summation appearing in Equation \eqref{eq:PCE} can be restricted to a subset of the $d$-dimensional multi-indices $\boldsymbol\alpha$. A standard truncation consists of polynomial orders up to a specified total degree equal to $p$. Other restrictions can be devised to promote parsimony or computational efficiency \citep{Tipireddy:2014}. In general, any such restruction can be expressed as
\begin{equation}
        \label{eq:truncated PCE}
        \mathbf Y = \mathcal{M}(\mathbf{X}) \approx \widehat{\mathbf Y} =\sum_{\boldsymbol\alpha\in\mathcal{A}} {\mathbf Y}_{\boldsymbol \alpha} \psi_{\boldsymbol\alpha}(\mathbf{X}),
\end{equation}
where $\mathcal{A}$ is a finite set of multi-indices. 
Clearly, given the orthogonality of the set $\psi_{\boldsymbol\alpha}(\mathbf{X})$, the PCE coefficients are readily expressed as
\begin{equation}
{\mathbf Y}_{\boldsymbol \alpha}  = \mathbb{E}\left[\mathbf Y\ \psi_{\boldsymbol\alpha}\right] = \int_{\mathbb{R}^d} \mathcal{M}(\mathbf x) \psi_{\boldsymbol\alpha}(\mathbf x) p_{\mathbf X}(\mathbf x) \ d\mathbf x \ .
\end{equation}
This last integral is over $\mathbb{R}^d$ and is often prohibitive to evaluate. Approximating it using various quadrature rules including Monte Carlo sampling and sparse quadrature is an active research area \cite{Tipireddy:2014}.  In the present work, and given the low-dimensional character of our problem $(d=4)$, we rely on least-square approach, which should yield sufficient accuracy with relatively few number of samples ~\citep{hadigol2018least,hampton2015coherence}.

\subsection{Joint PDF Characterization via PCE}
\begin{figure}[htb!]
	\centering
	\includegraphics[width=0.6\textwidth]{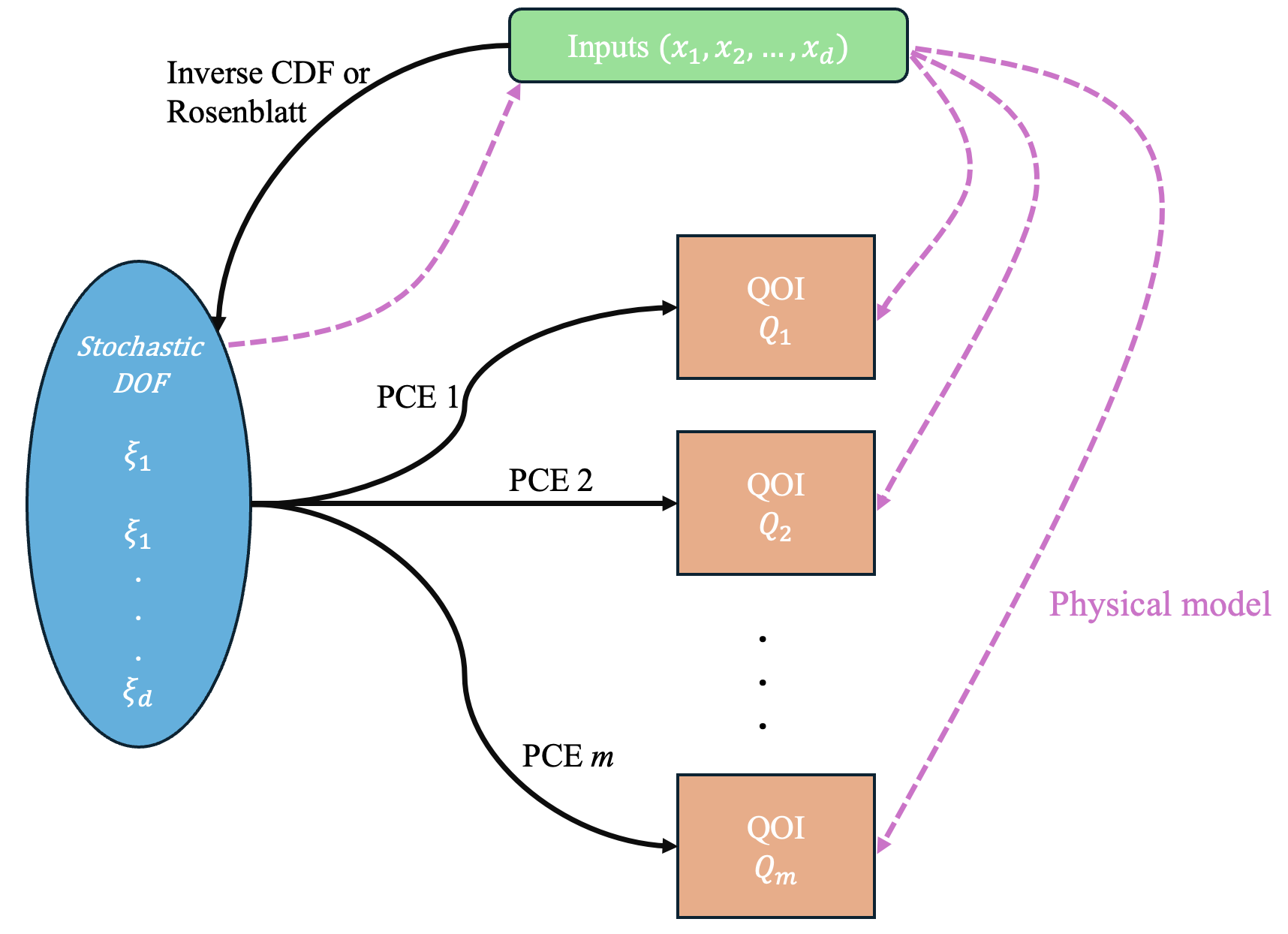}	
	\caption{The schematic diagram of the obtaining joint PDF via multiple PC expansions} 
	\label{fig: PCE joint PDF framework}
\end{figure}

The PC expansion provides an efficient way to understand and quantify the propagation of uncertainties from input parameters to quantities of interest in the physical system. The probability density of quantities of interest can be easily obtained by evaluating the PC expansion using Monte Carlo samples with minimum computational efforts. This procedure can be applied for each single quantity of interest and obtain its corresponding pdf. One useful but relatively under-explored advantage of PCE is that it can be used to characterize the joint distribution of quantities of interest under the same input uncertainty. Specifically, consider two quantities of interest denoted by random variables $\mathbf Y_1$ and $\mathbf Y_2$, respectively. The inherent uncertainties in the system are parameterized by the same random vector $\mathbf{X}$. We first convert the random vector $\mathbf{X}$ into multivariate standard Gaussian vector $\boldsymbol{\xi}$ using some transport map such as the inverse cumulative density function (CDF) or Rosenblatt transformation. Next, we construct the PC expansion $\mathbf{Y}_1=\sum_{\boldsymbol\alpha} \mathbf Y_{1,\boldsymbol{\alpha}}\psi_{\boldsymbol{\alpha}}(\boldsymbol{\xi})$ and $\mathbf Y_2=\sum_{\boldsymbol{\alpha}} \mathbf Y_{2,\boldsymbol{\alpha}}\psi_{\boldsymbol{\alpha}}(\boldsymbol{\xi})$ individually. Since both PC expansions are in terms of the same germs $\boldsymbol{\xi}$, one can sample the germs $\boldsymbol{\xi}$ and evaluate both PCEs at the same sample, resulting in samples of joint occurrence of $\mathbf{Y}_1$ and $\mathbf{Y}_2$. The joint distribution of the two quantities $\mathbf{Y}_1$ and $\mathbf{Y}_2$ can then be  accurately estimated from these samples. By tracing back to the uncertain sources, all quantities of interest can be expressed in the same chaos space and quantified in a unified framework. The schematic diagram is illustrated in Figure~\ref{fig: PCE joint PDF framework}. The input uncertain parameters $\mathbf{X} = (X_1,\cdots,X_d)$ of the system are first mapped to homogeneous chaos space, and uncertainties are represented by stochastic degrees of freedom (SDOF) with germs $\xi_1,\cdots,\xi_d$. Then, PC expansion from these SDOFs is constructed for each QOI individually. Joint distributions of the QOIs can be characterized by sampling these PCEs. Note that instead of building PCEs using available evaluations of physical models as described above, an alternative approach is to use these available joint samples of $Q_1,Q_2,\cdots,Q_m$ to characterize the joint distribution directly (for example, using kernel density estimation or other non-parametric approaches). However, this approach is typically less accurate, especially in the case where the dimension $m$ is high since high dimensional joint distributions requires many more samples to characterize. Our PCE approach can generate millions of samples with minimum computational efforts, paving the way to quantify high-dimensional joint distributions.

\section{Physical model and Computational setup}

The set of non-linear partial differential equations governing the detonative combustion is solved using density-based OpenFOAM solver for numerical simulations of compressible reactive flow and associated shock structures.
The flow turbulence is taken care of using large eddy simulation (LES) where kEqn-model~\citep{yoshizawa1986statistical, greenshields2023} is adopted.
The HLLC flux scheme with vanLeer limiters is applied to deal with convective terms. For time advancement, a second-order strong-stability-preserving scheme is adopted, and diffusion terms are solved using a Gauss scheme possessing second-order accuracy with corrected gradients.
The detailed chemical mechanism with 10 species and 25 elementary reactions by Marinov~\citep{marinov1996detailed} is used for chemical reactions in the simulations setup. This chemical mechanism predicts detonation process characteristics accurately~\citep{marcantoni2017rhocentralrffoam}. 
%
 
\begin{figure}[htbp!]
	\centering
	\includegraphics[width=0.7\textwidth]{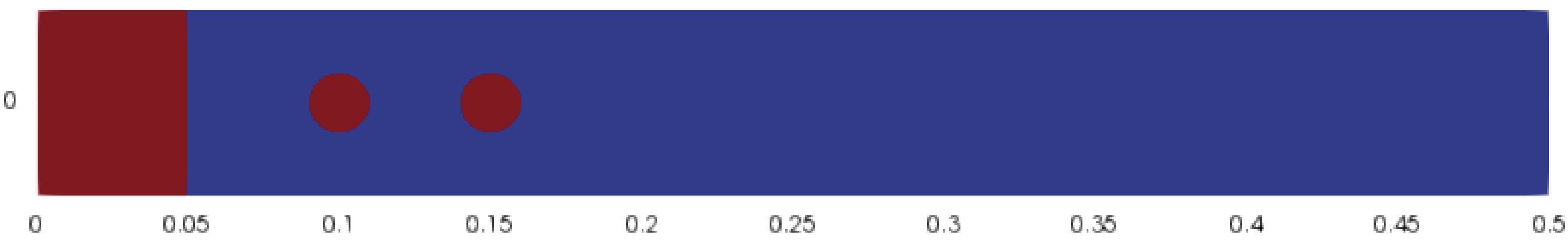}	
	\caption{The schematic diagram of the physical model} 
	\label{fig:physicalModel}%
\end{figure}

%
The propagation of detonation waves is simulated in a two-dimensional domain of size $500 \times 61.4 \text{ mm}^{2}$.
The schematic diagram of the physical model setup for detonation-driven shock interaction with $H_{2}-O_{2}$ fuel is in Figure~\ref{fig:physicalModel}.
Here, two circular-shaped fuel droplets are placed in the computational domain where fuel is a stoichiometric hydrogen-air mixture. Each droplet consists of a stoichiometric hydrogen–air mixture, while the rest of the domain is filled with ambient air, modeled as an $N_2$–$O_2$ mixture in a volumetric ratio of 0.78:0.22.
The initial conditions for the entire domain are set to atmospheric conditions, with a uniform temperature $T_0 = 300$ K and pressure $p_0 = 100$ kPa. To initiate  detonation, an ignition zone of dimensions $60 \times 61.4~\text{mm}^2$ is placed at the left end of the domain. Within this zone, the temperature and pressure are elevated to $10\,T_0$ and $50\,p_0$, respectively. The ignition region contains the same stoichiometric hydrogen–air mixture as the droplets.
Moreover, the channel's right end is given waveTramissive outflow boundary conditions to mitigate numerical reflection, while top and bottom boundaries are considered adiabatic with slip conditions.

%
%

%
 
\section{Results and discussions}

\begin{figure}
    \centering
    \begin{subfigure}{\textwidth}
        \centering
        \includegraphics[width=0.92\textwidth]{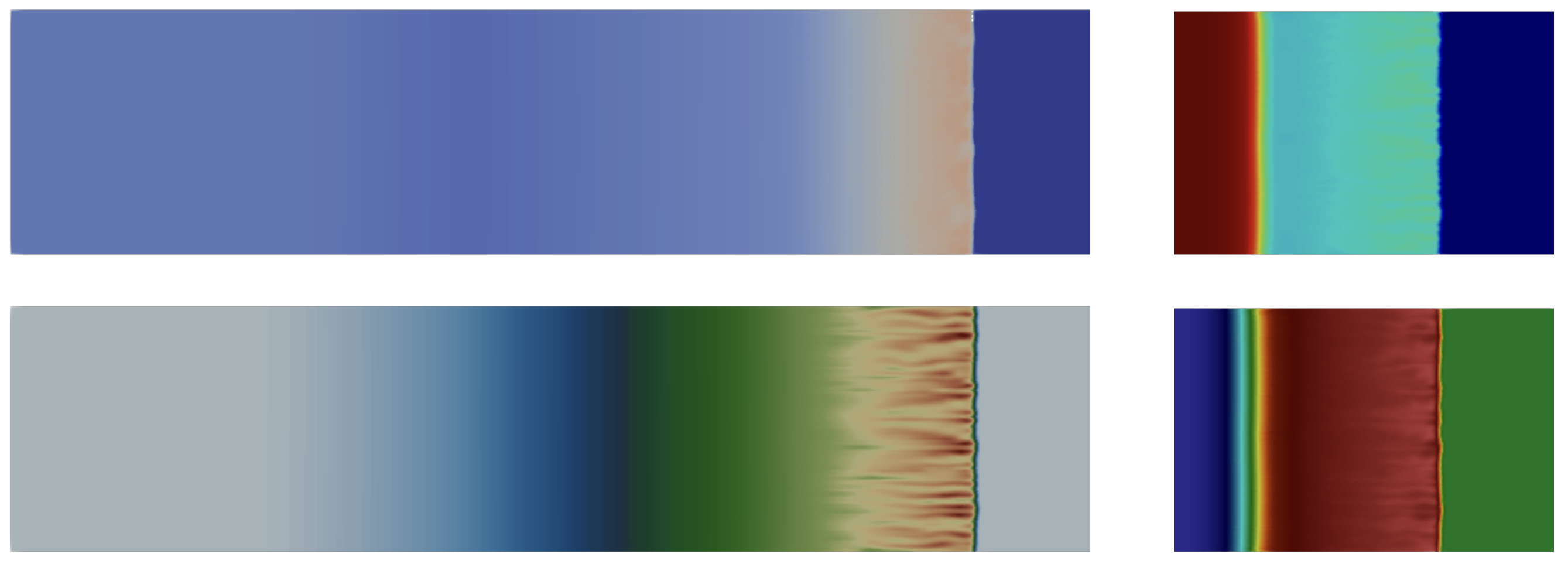}
        \subcaption{No droplets in the computational domain.}
        \label{fig: field-no drop}
    \end{subfigure}

    \begin{subfigure}{\textwidth}
        \centering
        \includegraphics[width=0.92\textwidth]{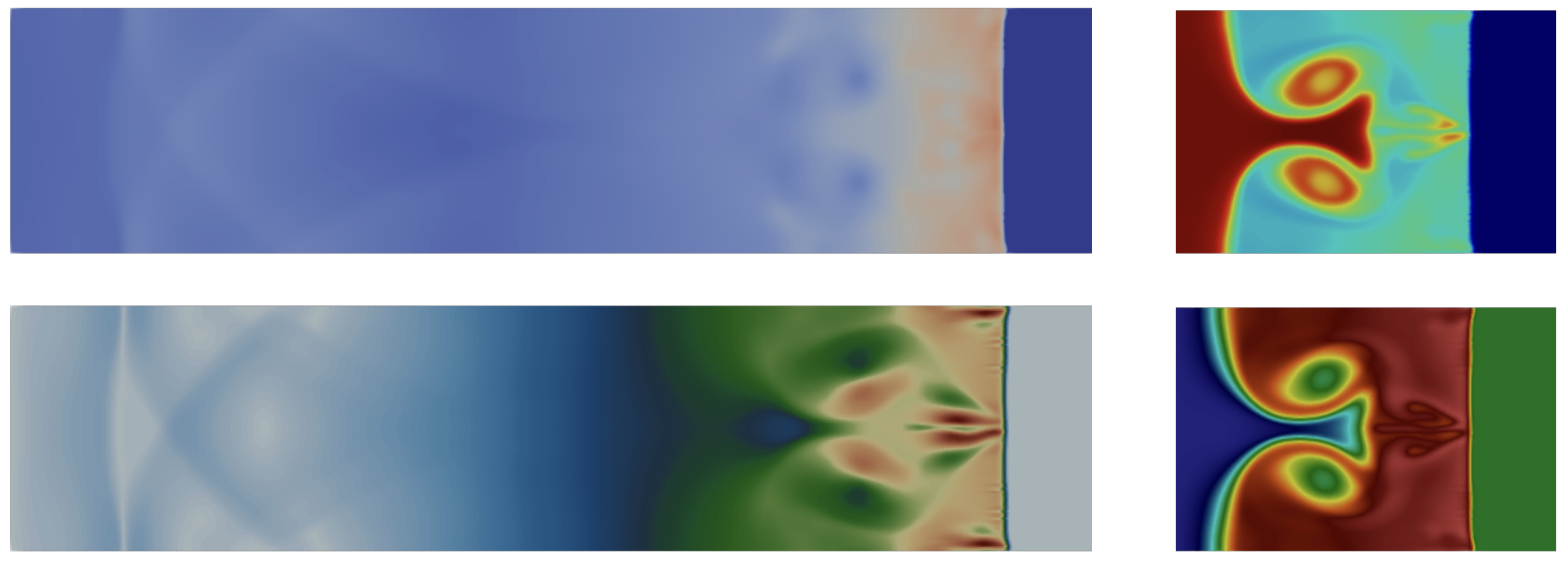}
        \subcaption{One droplet in the computational domain.}
        \label{fig: field-one drop}
    \end{subfigure}

    \begin{subfigure}{\textwidth}
        \centering
        \includegraphics[width=0.92\textwidth]{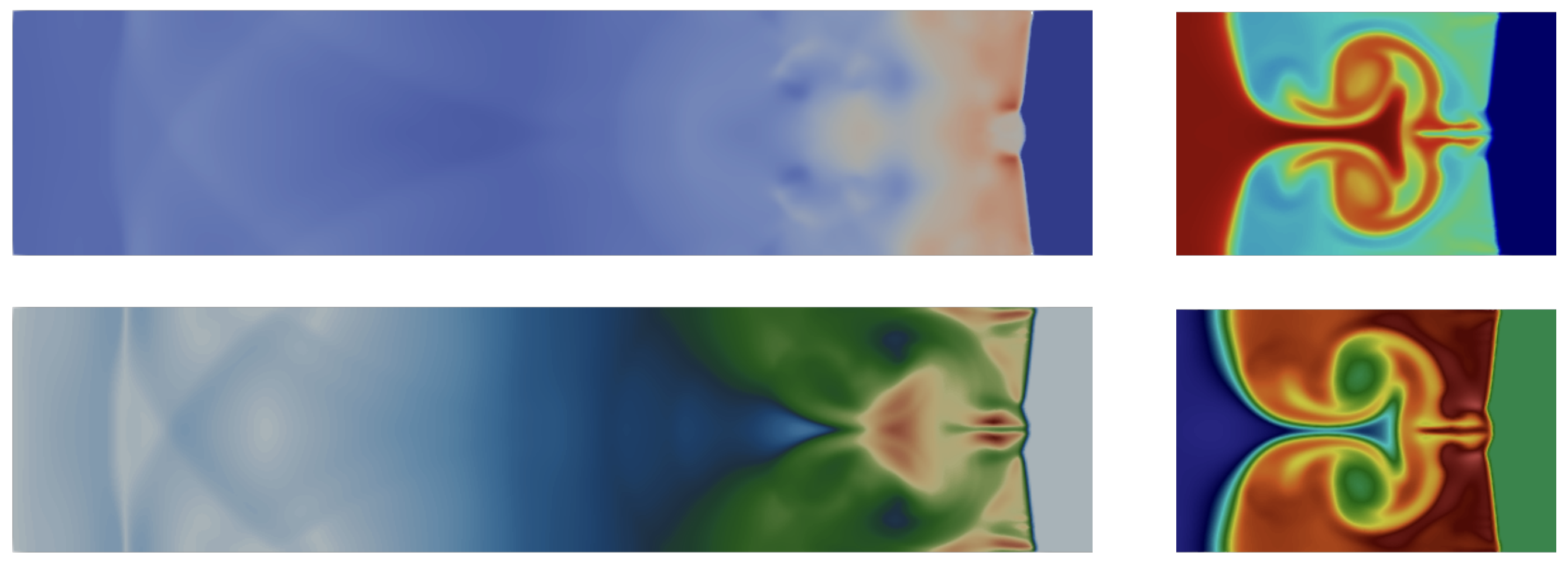}
        \subcaption{Two droplets in the computational domain.}
        \label{fig: field-two drop}
    \end{subfigure}
    
    \caption{Instantaneous contours of different physical quantities at three conditions at the simulation time step 32. For each subfigure, from left to right, top to bottom, the plots are pressure, temperature, velocity (in the horizontal direction), and density fields, respectively. For the temperature and density fields, only the rightmost third of the computational domain is plotted because the other parts are nearly quiescent.}
    \label{fig: Instantaneous contours for PTVD}
\end{figure}

\begin{figure}
    \centering
    \begin{subfigure}{\textwidth}
        \centering
        \includegraphics[width=0.92\textwidth]{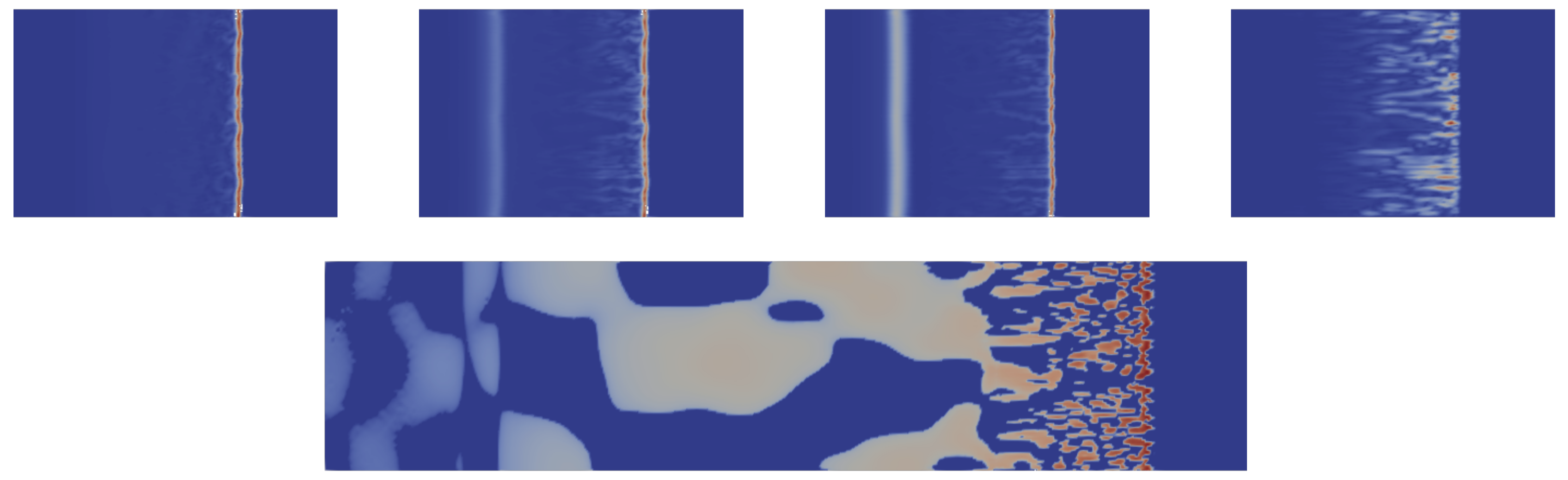}
        \subcaption{No droplets in the computational domain.}
        \label{fig: grad field-no drop}
    \end{subfigure}

    \begin{subfigure}{\textwidth}
        \centering
        \includegraphics[width=0.92\textwidth]{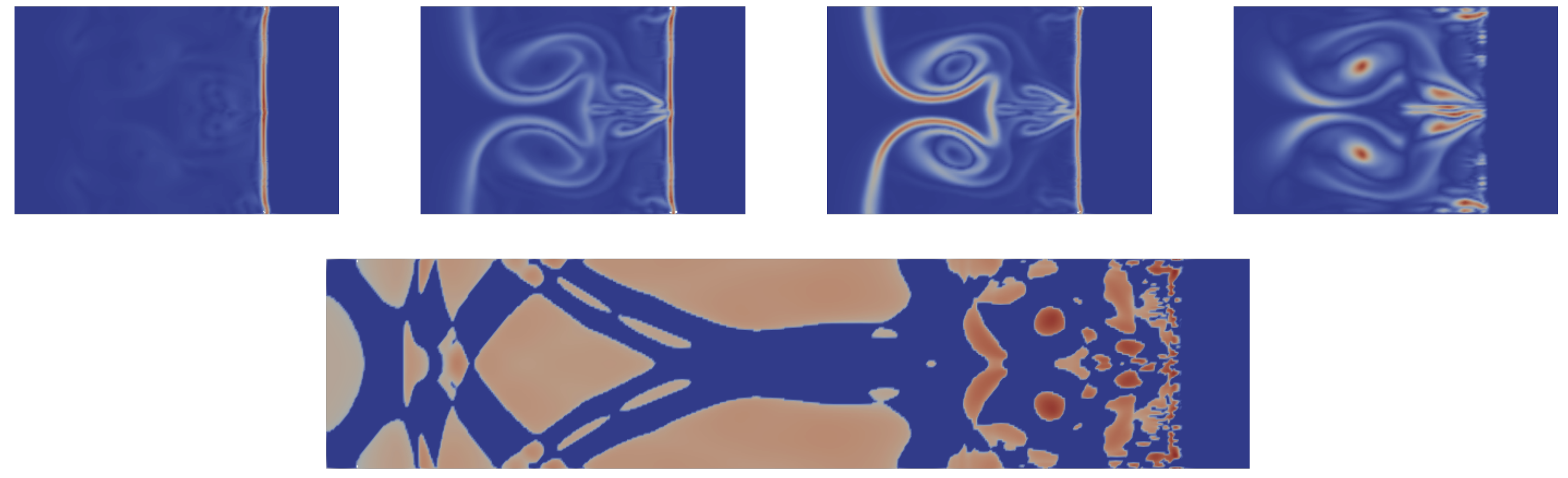}
        \subcaption{One droplet in the computational domain.}
        \label{fig: grad field-one drop}
    \end{subfigure}

    \begin{subfigure}{\textwidth}
        \centering
        \includegraphics[width=0.92\textwidth]{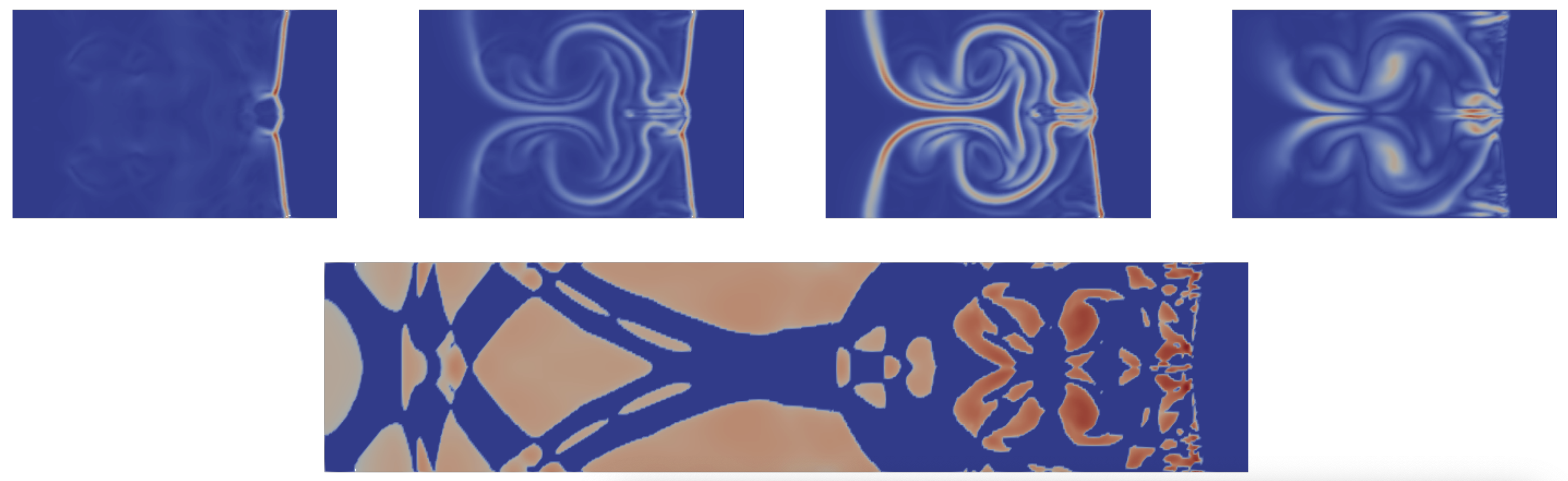}
        \subcaption{Two droplets in the computational domain.}
        \label{fig: grad field-two drop}
    \end{subfigure}
    
    \caption{Instantaneous contours of gradient fields with different number of droplets in the channel at the simlation time step 32. For each subfigure, from left to right, top to bottom, the plots are the magnitudes of pressure gradients, density gradients, temperature gradients, vorticity, and Q-criterion fields, respectively. For all fields in the first row, only the rightmost third of the computational domain is plotted due to the observed unique patterns in this area, while the rest of the domain are nearly quiescent. In the second row, the whole Q-criterion field is plotted.}
    \label{fig: Instantaneous gradient contours for PTVD}
\end{figure}

Figure~\ref{fig: Instantaneous contours for PTVD} presents the instantaneous flow field for the no-droplet, one-droplet, and two-droplet cases at the simulation time step 32. For each case, we display the contours of pressure, temperature, velocity in the horizontal direction, and density (PTURho contours), arranged sequentially from left to right and top to bottom. For the panels in the right columns, only the rightmost third of the channel is shown, as the upstream regions are nearly homogeneous and do not exhibit significant spatial variations. Complementarily, Figure~\ref{fig: Instantaneous gradient contours for PTVD} depicts the corresponding gradient-based fields: contours of pressure gradient ($\nabla p$), density gradient ($\nabla \rho$), temperature gradient ($\nabla T$), vorticity, and $Q$-criterion (collectively referred to as $\nabla$PTR contours). As with the PTURho plots, the top row panels in this figure are restricted to the rightmost third of the channel to emphasize the regions of strongest gradients and flow structures.

Figure~\ref{fig: field-no drop} illustrates the instantaneous flow field for the case without fuel droplets. A detonation wave is initiated by the driver section located at the left boundary of the shock tube and propagates toward the right. The wavefront exhibits sharp discontinuities. Notably, the detonation front forms a thick, planar slab structure, with its leading front layer (FLDF) approximately parallel to the rear layer (RLDF), indicative of a quasi-one-dimensional, symmetric propagation.
Figure~\ref{fig: grad field-no drop} presents the corresponding gradient-based field. The pressure gradient magnitude ($\nabla p$) is strongly localized at the FLDF, reflecting the abrupt pressure rise across the shock. The temperature and density gradients ($\nabla T$, $\nabla \rho$) are also significant at the FLDF but persist, albeit at reduced magnitude, across the RLDF. Vorticity is sharply concentrated at the FLDF and rapidly diminishes before reaching the RLDF, consistent with the rapid dissipation of rotational motion downstream. These high gradient magnitudes indicate sharp spatial variations in thermodynamic variables, intrinsic to the detonation structure.

When a single fuel droplet is introduced into the domain, two key phenomena emerge, as illustrated in the PTURho and $\nabla$PTR contour plots in Figures~\ref{fig: field-one drop} and~\ref{fig: grad field-one drop}, respectively. First, the interaction between the propagating detonation front and the droplet generates complex flow structures and perturbs the otherwise planar detonation front, leading to localized inhomogeneities in thermodynamic fields. Second, a secondary detonation wave is observed propagating in the opposite direction. 
The PTURho contours reveal the formation of two symmetric eddies in temperature and density fields near the FLDF, aligned along the channel centerline ($y = 0$).  Additionally, pressure and velocity distributions show the evolution of the detonation cell and its movements in a backward direction. The leftward-moving detonative shock looks significant, but it is not as strong as the right-moving detonative shock. Strong gradients are still observed at FLDH. However, unlike the no-droplet case, the RLDF is no longer distinguishable; instead, strong thermal and density gradients are sustained between the main rotating eddies. Smaller vortices also emerge near the FLDF.
Moreover, one can observe that the FLDF is slightly ahead of the case with no droplet, representing the quantitative growth of detonation-front speed. 

When a second droplet is introduced, the interaction becomes more intense, as expected, as shown in Figure~\ref{fig: field-two drop}. The pressure contours indicate the formation of multiple backward-propagating waves. Velocity contours get higher magnitudes. The temperature and density fields develop a distinct bow-shaped structure surrounding the eddies generated by the first droplet interaction. This bow is symmetrically aligned about the channel centerline and flanked by two low-intensity vortices near the top and bottom walls. Strong temperature and density gradients are observed along the outer boundary of this bow-shaped layer, highlighting the amplification of thermal and fluid-dynamic effects due to the second droplet.
These results suggest that the presence of the second droplet induces additive effects across all physical fields, intensifying the overall flow dynamics.

Importantly, the pressure gradient contours ($\nabla p$) in Figure~\ref{fig: Instantaneous gradient contours for PTVD} show minimal change between the no-droplet and one-droplet cases—both present as vertically aligned, straight gradient bands centered along the $y$-axis. However, in the two-droplet case, this structure is significantly altered. In the upper half of the domain, the vertical line is deflected and forms a slanted arc, with a semi-circular window opening near the centerline ($y = 0$); a mirrored structure appears in the lower half. These perturbations result from intensified detonative forces caused by the additional fuel distributed near the domain center, demonstrating nonlinear interactions and amplification effects due to the second droplet.

Additionally, the $\nabla \rho$, $\nabla T$, and vorticity contours exhibit distinct morphological changes. In the one-droplet case, their distribution resembles the horizontal cross-section of a flying bird having a face attached to FLDF. In the two-droplet case, these contour curves resemble the behavior of such birds expanding their wings to gain acceleration. This can be seen by patterns, e.g., contour curves near the center-line of $y$-axis shrink (face to tail or body of the bird), and wings expand. A very important behavior to notice is that higher-density contours have higher thermal gradients. This in turn causes extreme velocity gradients, which results in higher levels of vorticity at the low-density site. 
Finally, the $Q$-criterion contours delineate regions dominated by rotational motion, clearly outlining the detonation front and capturing the structure of the backward-propagating detonation cell. These results collectively highlight the complex, multi-scale flow structures that arise from fuel-droplet interactions in detonation environments.

The interaction of a detonative shock wave with two fuel droplets is highly sensitive to several parameters. These include: (1) variations in distance between droplets, (2) the vertical misalignment of droplets relative to the channel centerline, (3) the fuel composition within each droplet, and (4) the droplet size. The following sections provide a systematic investigation of these factors, analyzing their influence on detonation wave dynamics using key flow diagnostics: (a) pressure gradient, (b) temperature gradient, (c) density gradient, (d) vorticity, and (e) $Q$-criterion.

\subsection{Effect of droplet distance}
\begin{figure}[htbp!]
    \centering
    \begin{subfigure}{\textwidth}
        \centering
        \includegraphics[width=0.9\textwidth]{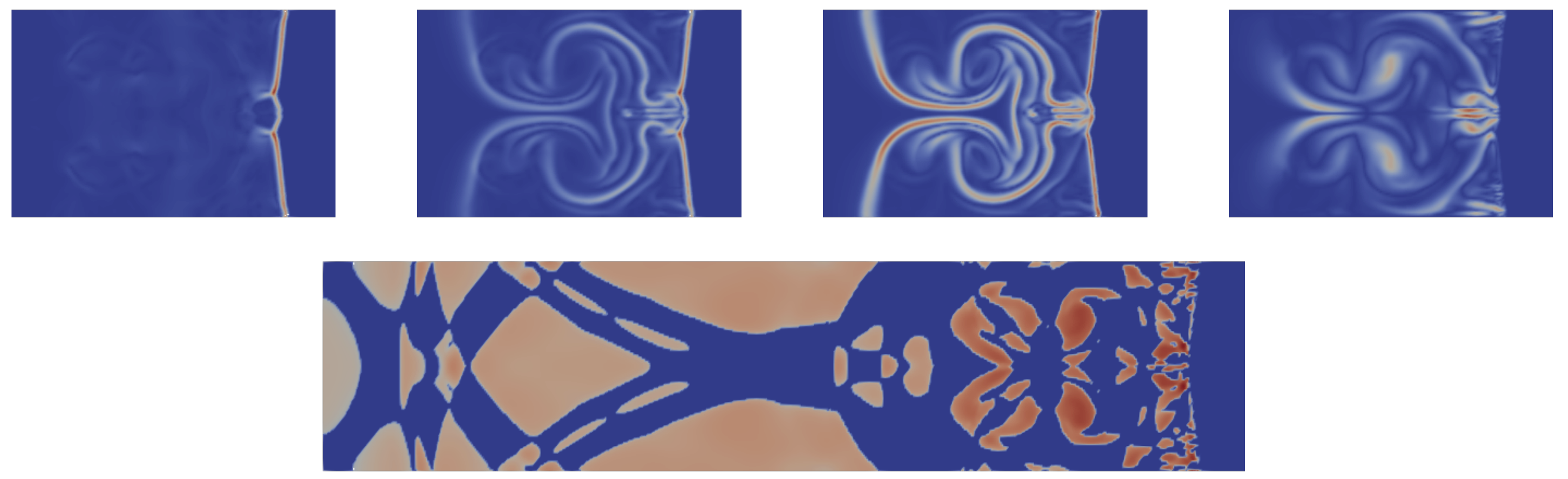}
        \caption{Droplet distance is 50 mm.}
        \label{fig: grad field-dist 1}
    \end{subfigure}

    \begin{subfigure}{\textwidth}
        \centering
        \includegraphics[width=0.9\textwidth]{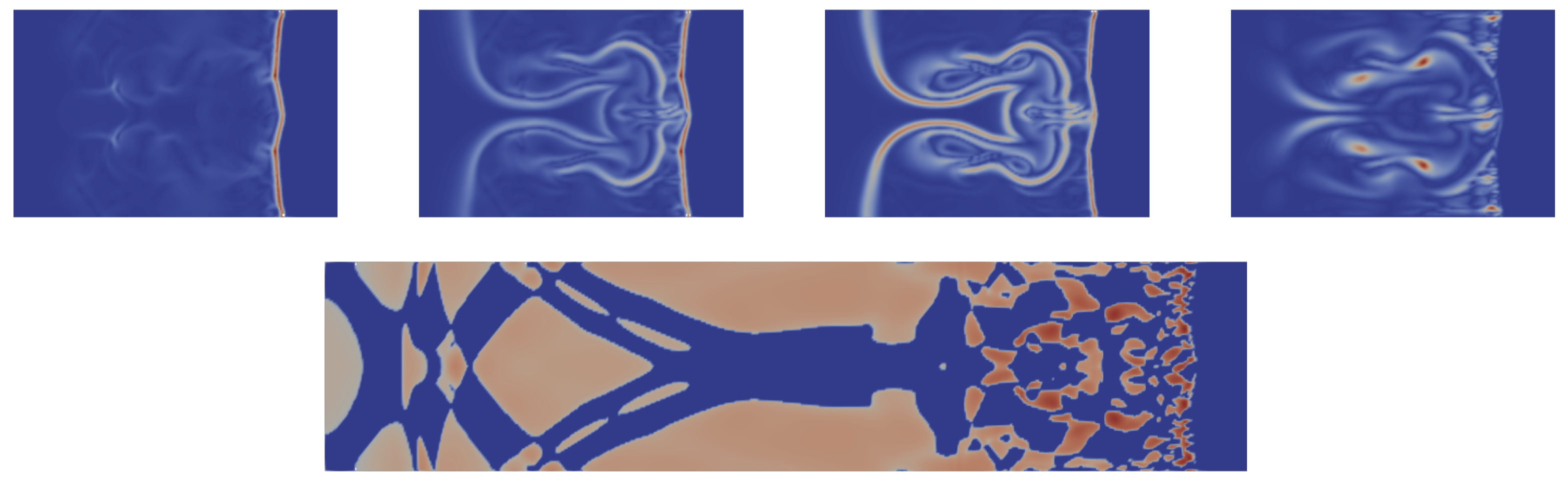}
        \caption{Droplet distance is 75 mm.}
        \label{fig: grad field-dist 2}
    \end{subfigure}

    \begin{subfigure}{\textwidth}
        \centering
        \includegraphics[width=0.9\textwidth]{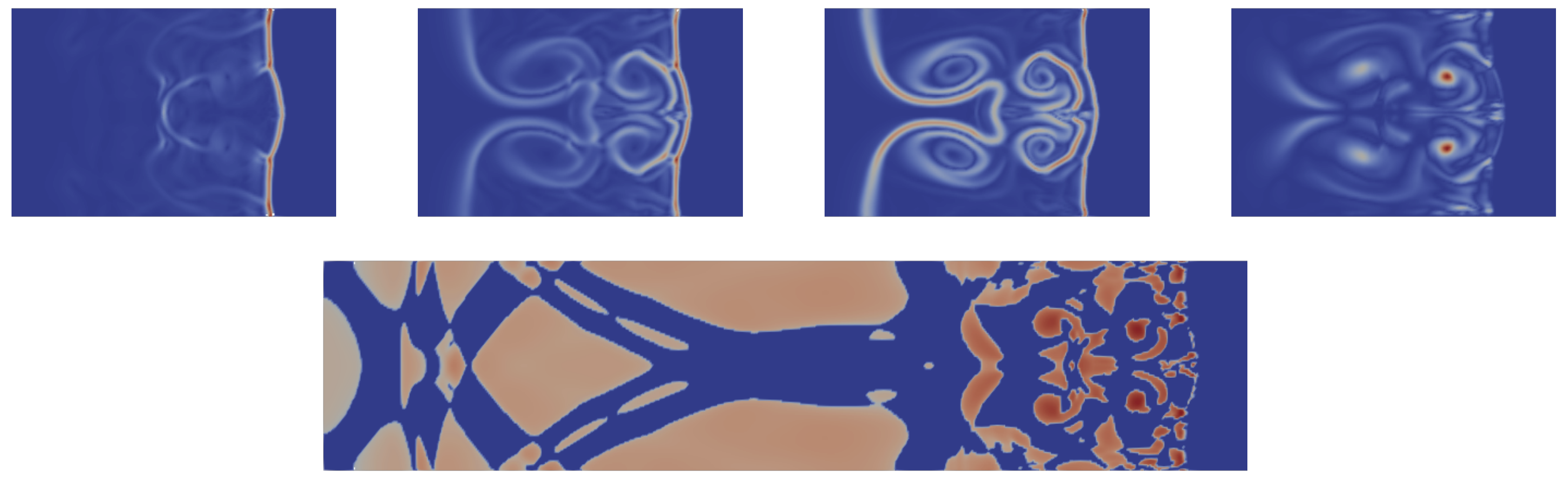}
        \caption{Droplet distance is 100 mm.}
        \label{fig: grad field-dist 3}
    \end{subfigure}
    
    \caption{Instantaneous contours of gradient fields with different distances between droplets at the simulation time step 32. For each subfigure, from left to right, top to bottom, the plots are the magnitudes of pressure gradients, density gradients, temperature gradients, vorticity, and Q-criterion fields, respectively. For all fields in the first row, only the rightmost 1/3 of the computational domain is plotted due to the observed unique patterns in this area, while the rest of the domain are near quiescent. In the second row, the whole Q-criterion field is plotted.}
    \label{fig: Instantaneous gradient contours for PTVD with different droplet distances.}
\end{figure}

In this analysis, the first droplet is placed 100 mm from the left boundary of the domain. The second droplet is placed at three different streamwise distances relative to the first: 50 mm, 75 mm, and 100 mm. The corresponding $\nabla$PTR fields are shown in Figure~\ref{fig: Instantaneous gradient contours for PTVD with different droplet distances.}.

When droplets are symmetrically located with respect to the centerline of the channel, the flow field exhibits mirror-image patterns. As the inter-droplet distance increases, the spatial extent of the $\nabla \rho$, $\nabla T$, and vorticity contours broadens horizontally. For the 100 mm spacing case, rather than a single pair of strong interacting eddies, two distinct doublet eddy structures emerge, indicating reduced coupling between the droplets.
These observations imply that increasing the horizontal separation between droplets attenuates the nonlinear interactions and additive reinforcement of the detonation structure. Furthermore, the detonation front (FLDF) position progressively lags as the distance increases, signifying a weakening of the detonation strength due to diminished localized energy deposition. This highlights the critical importance of droplet placement in practical detonation engine configurations to ensure sustained and robust combustion.

\subsection{Effect of vertical alignment}
The cases discussed so far assumed symmetric droplet placement across the centerline, which naturally leads to symmetric flow structures. Here, we examine how vertical misalignment of the second droplet disrupts this symmetry.

\begin{figure}[htbp!]
    \centering
    \begin{subfigure}{\textwidth}
        \centering
        \includegraphics[width=0.9\textwidth]{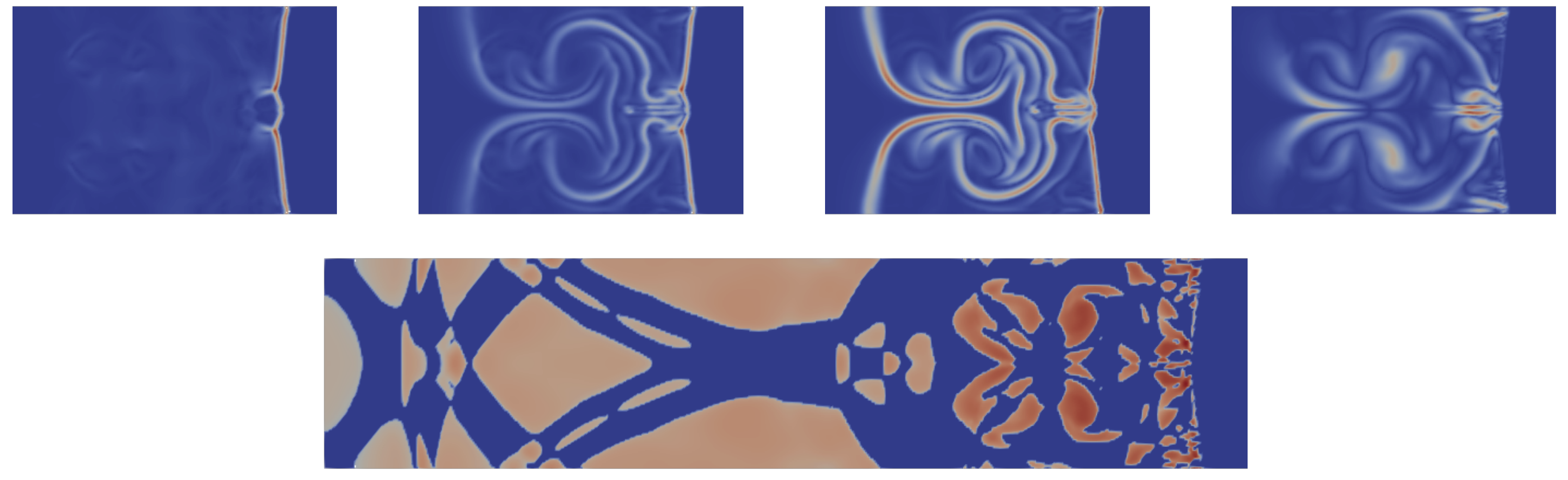}
        \caption{Droplet alignment angle 0$^\circ$.}
        \label{fig: grad field-angle 1}
    \end{subfigure}

    \begin{subfigure}{\textwidth}
        \centering
        \includegraphics[width=0.9\textwidth]{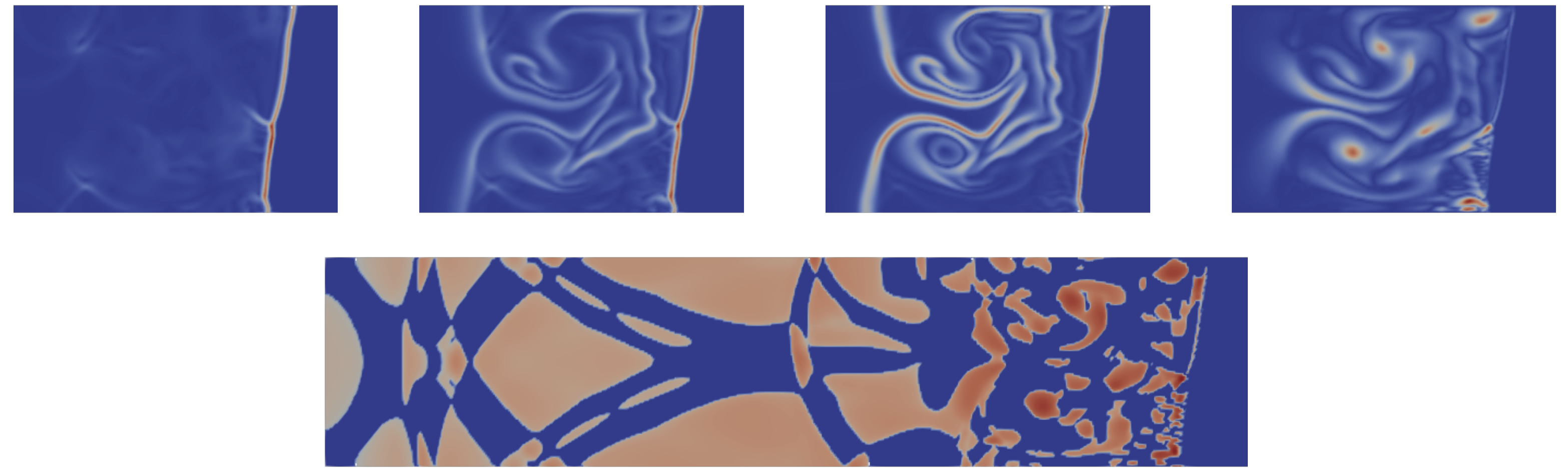}
        \caption{Droplet alignment angle 5$^\circ$.}
        \label{fig: grad field-angle 2}
    \end{subfigure}

    \begin{subfigure}{\textwidth}
        \centering
        \includegraphics[width=0.9\textwidth]{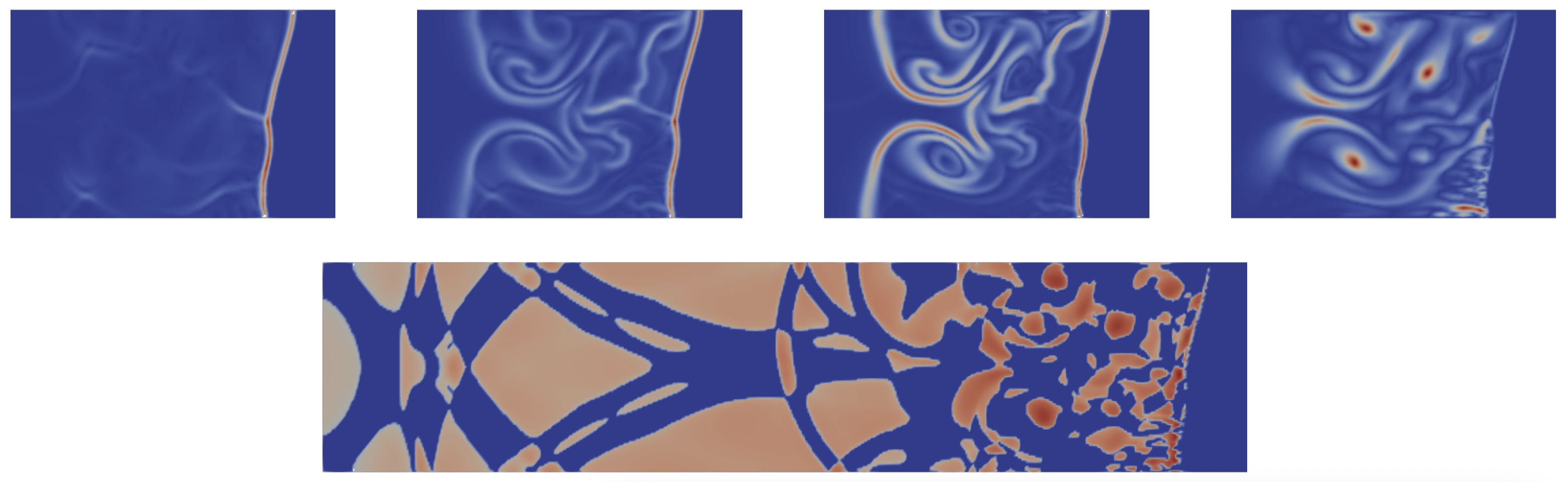}
        \caption{Droplet alignment angle 10$^\circ$.}
        \label{fig: grad field-angle 3}
    \end{subfigure}
    
    \caption{Instantaneous contours of gradient fields with different alignment angles between droplets at the simulation time step 32. For each subfigure, from left to right, top to bottom, the plots are the magnitudes of pressure gradients, density gradients, temperature gradients, vorticity, and Q-criterion fields, respectively. For all fields in the first row, only the rightmost 1/3 of the computational domain is plotted due to the observed unique patterns in this area, while the rest of the domain are nearly quiescent. In the second row, the whole Q-criterion field is plotted.}
    \label{fig: Instantaneous gradient contours for PTVD angle effect.}
\end{figure}

In this study, the first droplet remains fixed at $y = 0$, while the second droplet is placed 50 mm downstream but at increasing vertical angles (0$^\circ$, 5$^\circ$, and 10$^\circ$ relative to the centerline). The resulting $\nabla$PTR fields are shown in Figure~\ref{fig: Instantaneous gradient contours for PTVD angle effect.}.
In the 0$^\circ$ case, the $\nabla$PTR contours remain symmetric, reflecting balanced interactions. However, as the vertical offset increases, symmetry is progressively broken. The contours become distorted, with increased complexity and turbulence, particularly in regions of high gradient magnitude.
A notable trend is the emergence of a vertical component in all gradient fields, which introduces shear and delays the forward detonation front along the $x$-axis. This retardation effect diminishes the forward-propagating detonation strength, increasing the likelihood of incomplete combustion or detonation quenching. These findings underscore the symmetry effects in droplet injection strategies, as misalignment can degrade engine performance and lead to conditions associated with detonation unstart or instability.

\subsection{Effect of fuel concentration}
To assess the effect of fuel composition, three different mixtures are investigated: (1) pure hydrogen ($H_2$), (2) a binary hydrogen-oxygen mixture ($H_2$–$O_2$), and (3) a ternary hydrogen-oxygen-nitrogen mixture ($H_2$–$O_2$–$N_2$). The droplet positions and alignment remain fixed as in the symmetric 50 mm inter-droplet case.

\begin{figure}[htbp!]
    \centering
    \begin{subfigure}{\textwidth}
        \centering
        \includegraphics[width=0.9\textwidth]{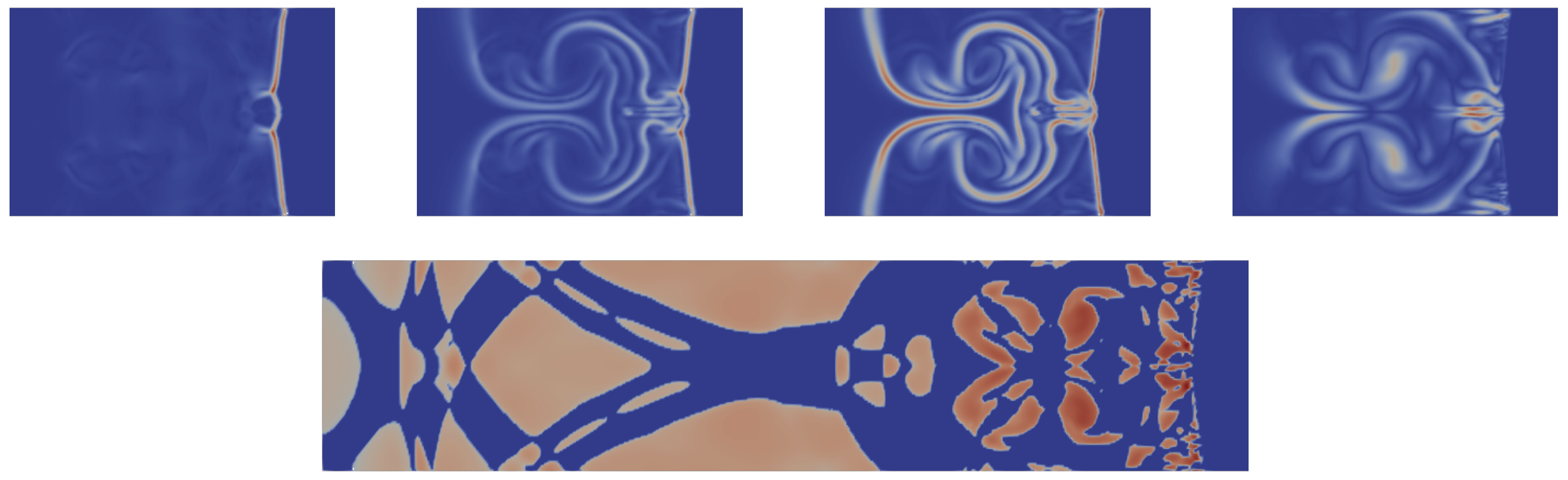}
        \caption{$H_2$ only.}
        \label{fig: grad field-O2 1}
    \end{subfigure}

    \begin{subfigure}{\textwidth}
        \centering
        \includegraphics[width=0.9\textwidth]{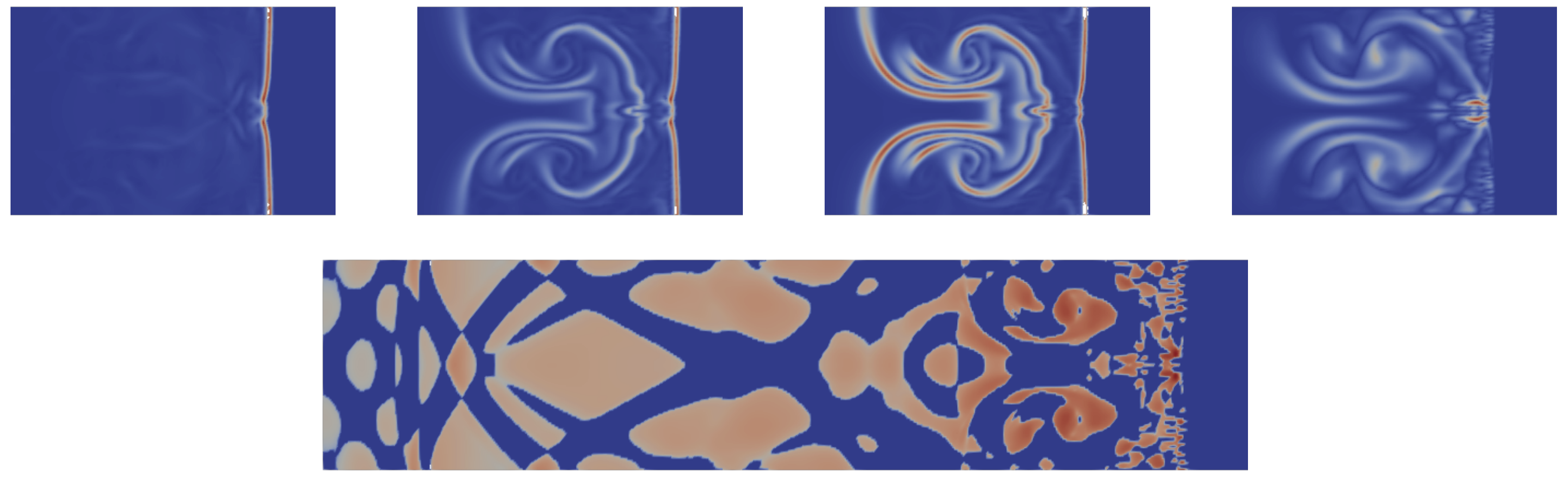}
        \caption{$H_{2}-O_{2}$ mixture.}
        \label{fig: grad field-O2 2}
    \end{subfigure}

    \begin{subfigure}{\textwidth}
        \centering
        \includegraphics[width=0.9\textwidth]{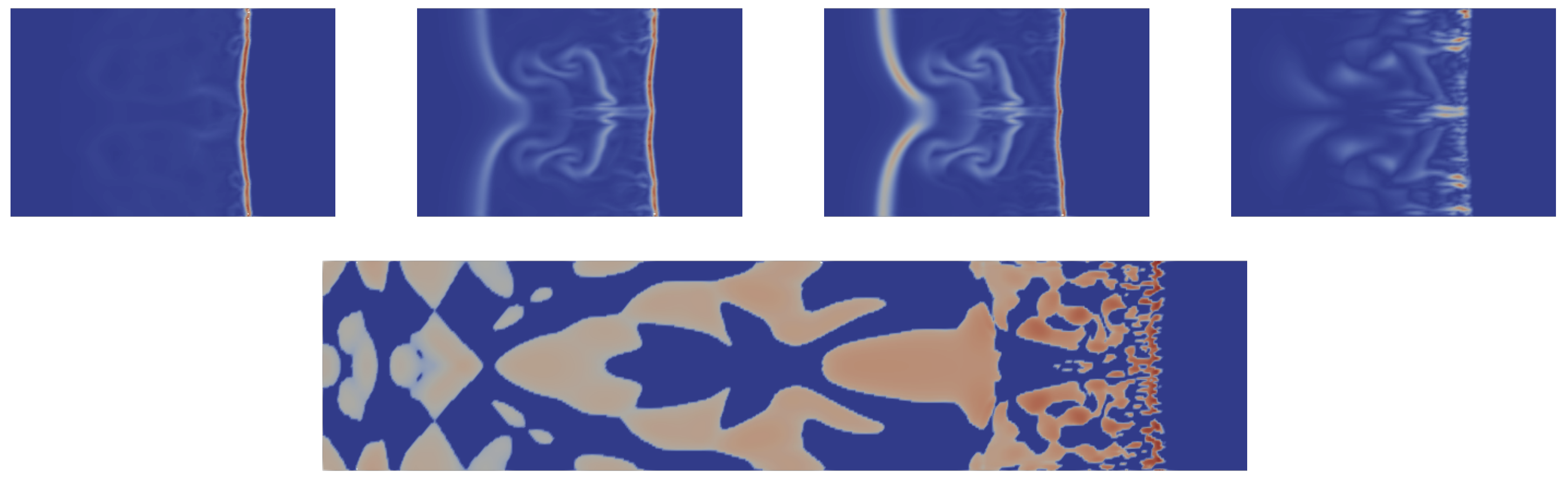}
        \caption{$H_{2}-O_{2}-N_{2}$ mixture.}
        \label{fig: grad field-O2 3}
    \end{subfigure}
    
    \caption{Instantaneous contours of gradient fields with different fuel compositions in droplets at the simulation time step 32. For each subfigure, from left to right, top to bottom, the plots are the magnitudes of pressure gradients, density gradients, temperature gradients, vorticity, and Q-criterion fields, respectively. For all fields in the first row, only the rightmost 1/3 of the computational domain is plotted due to the observed unique patterns in this area, while the rest of the domain are nearly quiescent. In the second row, the whole Q-criterion field is plotted.}
    \label{fig: Instantaneous gradient contours for PTVD fuel concen}
\end{figure}

Figure~\ref{fig: Instantaneous gradient contours for PTVD fuel concen} present the gradient fields for the three fuel types. As expected, higher hydrogen concentrations correspond to stronger pressure, temperature, and density gradients. The pure $H_2$ case produces the most intense gradients, with sharply defined detonation fronts and strong vorticity. As oxygen and nitrogen are added, the overall combustion intensity decreases.
In particular, the inclusion of nitrogen acts as a thermal diluent, absorbing energy and weakening the reaction zone. This results in smaller, weaker eddies and broader, less defined detonation fronts. These findings confirm the critical role of fuel reactivity in governing the strength and structure of the detonation wave.

\subsection{Effect of combined variations}
\begin{figure}[htbp!]
    \centering
    \begin{subfigure}{\textwidth}
        \centering
        \includegraphics[width=0.9\textwidth]{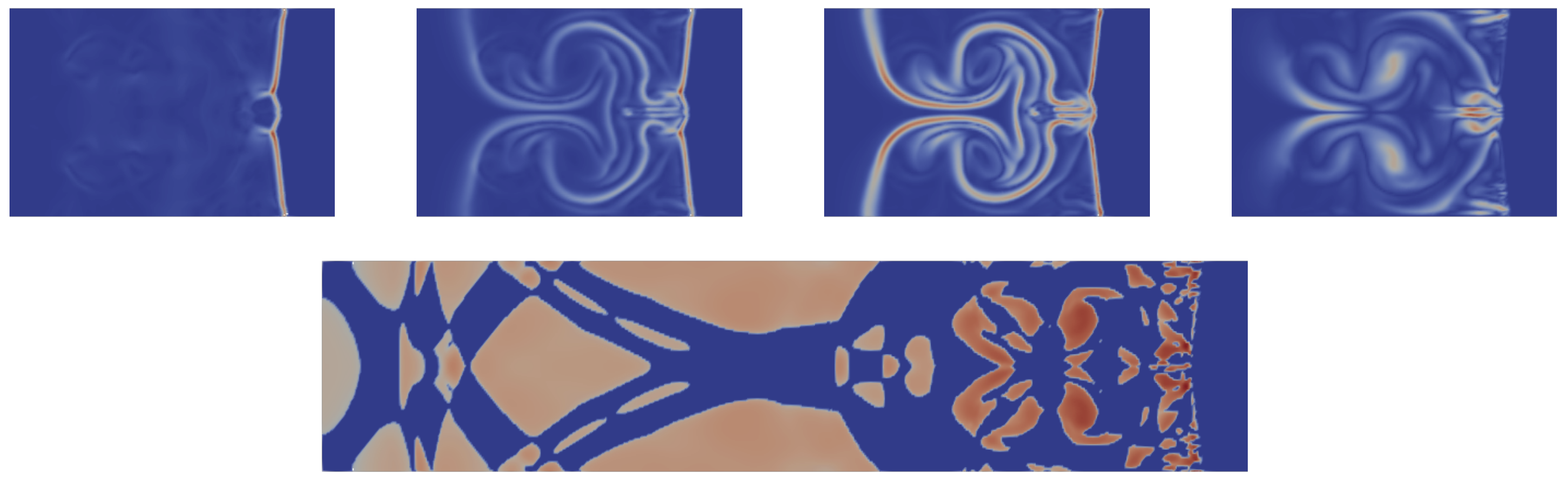}
        \caption{Case 1: $H_2$; Distance: 50 mm; Alignment angle: 0$^\circ$.}
        \label{fig: grad field-combine 1}
    \end{subfigure}

    \begin{subfigure}{\textwidth}
        \centering
        \includegraphics[width=0.9\textwidth]{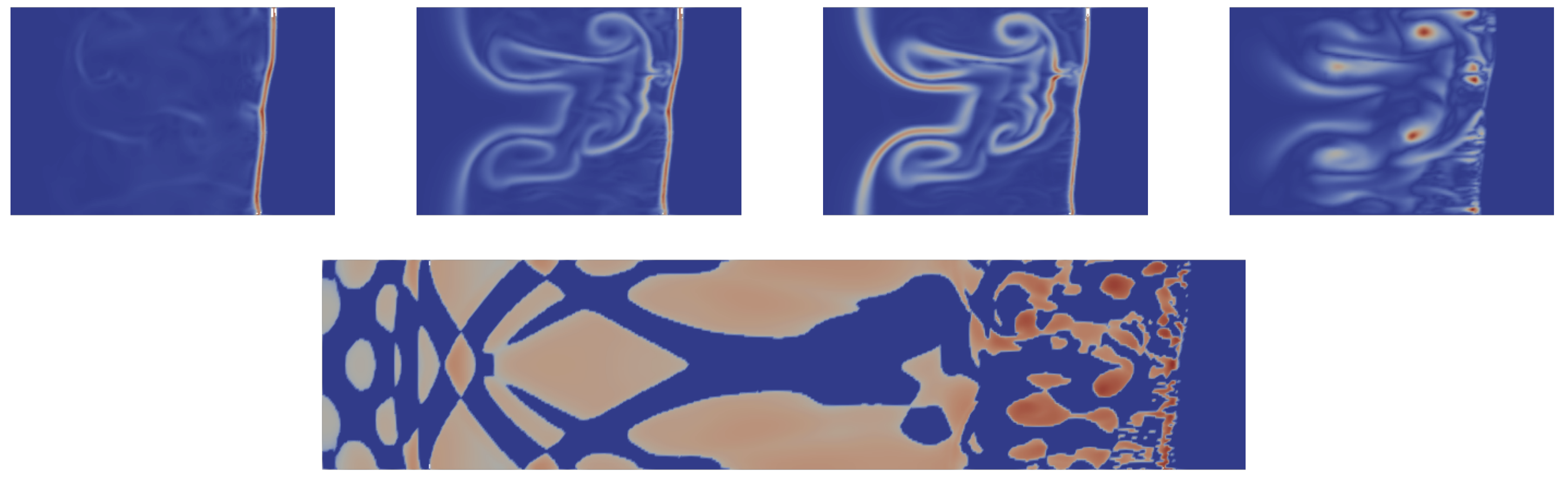}
        \caption{Case 2: $H_2-O_2$; Distance: 75 mm; Alignment angle: 5$^\circ$.}
        \label{fig: grad field-combine 2}
    \end{subfigure}

    \begin{subfigure}{\textwidth}
        \centering
        \includegraphics[width=0.9\textwidth]{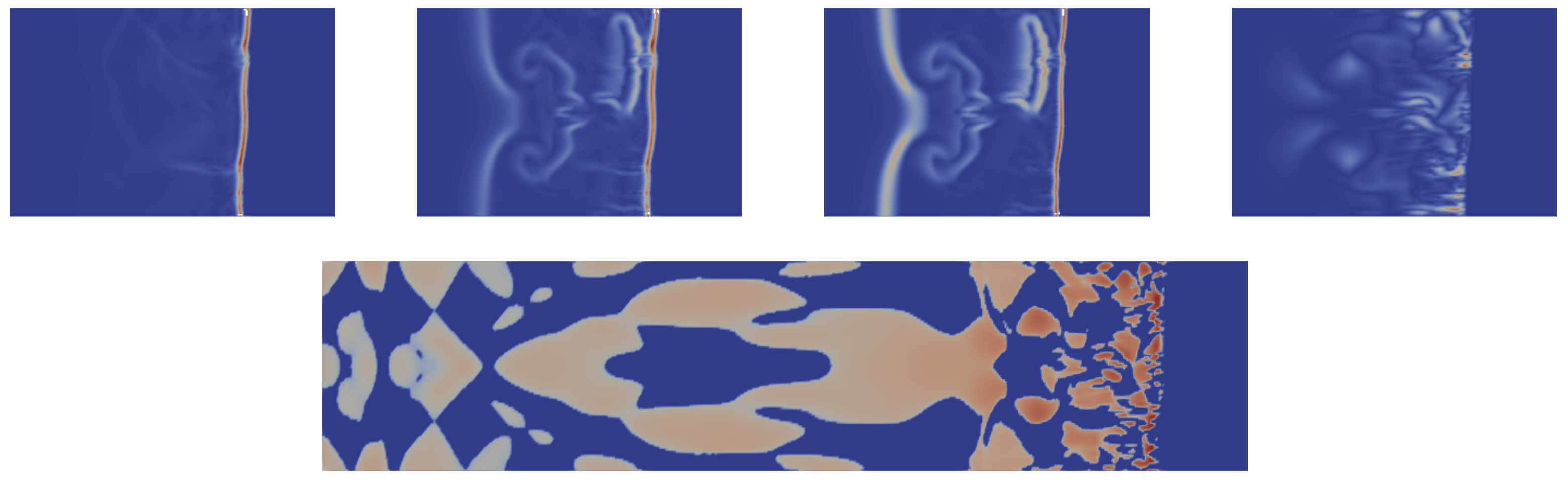}
        \caption{Case 3: $H_2-O_2-N_2$; Distance: 100 mm; Alignment angle: 10$^\circ$.}
        \label{fig: grad field-combine 3}
    \end{subfigure}
    
    \caption{Instantaneous contours of gradient fields with combined effects at the simulation time step 32. For each subfigure, from left to right, top to bottom, the plots are the magnitudes of pressure gradients, density gradients, temperature gradients, vorticity, and Q-criterion fields, respectively. For all fields in the first row, only the rightmost 1/3 of the computational domain is plotted due to the observed unique patterns in this area, while the rest of the domain are nearly quiescent. In the second row, the whole Q-criterion field is plotted..}
    \label{fig: Instantaneous gradient contours for PTVD combin}
\end{figure}

Finally, we explore the combined effect of varying all parameters—droplet spacing, alignment, fuel composition, and droplet size—simultaneously. The results are illustrated in Figure~\ref{fig: Instantaneous gradient contours for PTVD combin}.

The interplay of these parameters produces highly complex and detached gradient structures. Notably, the extent of horizontal asymmetry is most pronounced in high $H_2$-concentration cases, where strong gradients drive the formation of intense vortical features. As $H_2$ concentration is reduced—either by dilution with $O_2$/$N_2$ or by decreasing droplet size—the gradient intensity weakens, and the structures become more diffuse.

These results emphasize that optimizing detonation engine performance requires careful co-design of fuel composition, droplet placement, and injection configuration. Even small deviations in alignment or spacing can have significant effects on wave propagation and combustion efficiency.

%
%
%
%


\section{Uncertainty quantification results}
Besides the qualitative comparison of flow patterns to understand the effect of input parameters on detonation structures, this section provides quantitative results; more specifically, this section studies the propagation of uncertainties within this highly complicated system using polynomial chaos expansions. Based on previous discussions, we identified the important parameters that affect the detonation performance and viewed them as random input variables. They are $H2$ and $O2$ volume fraction in the droplet, droplet distance, droplet alignment angle, and droplet size. Since the volume faction of $H2$ and $O2$ sums to 1, one can drop one stochastic DoF and, in total, obtain four random parameters. Each random parameter is assumed to be independently Beta-distributed with shape parameters $(6,6)$. The ranges of Beta distributions for input parameters are shown in Table \ref{tab: beta parameters}. The reason for choosing beta distribution is that it is bounded and has a flexible density shape controlled by the shape parameters. 1000 samples are generated by the beta distributions and used as input for the simulations. These simulations provide the data to build PCEs in the following work. 

\begin{table}[htb]
\centering
\begin{tabular}{lcccr} 
\toprule
\textbf{Beta parameters} & \textbf{$O2$ Volume fraction} & \textbf{Distance} & \textbf{Angle} & \textbf{Droplet size} \\ \midrule
Shape parameters  & (6,6)  & (6,6)  & (6,6)  & (6,6)  \\
Lower limit  & 0.1  & 0.15  & 0  & 0.002  \\
Upper limit  & 0.9  & 0.2  & 0.02  & 0.01  \\ \bottomrule
\end{tabular}
\caption{Parameters of Beta distributions for all random variables.}
\label{tab: beta parameters}
\end{table}

\begin{figure}[htb]
    \centering
    \includegraphics[width=0.7\linewidth]{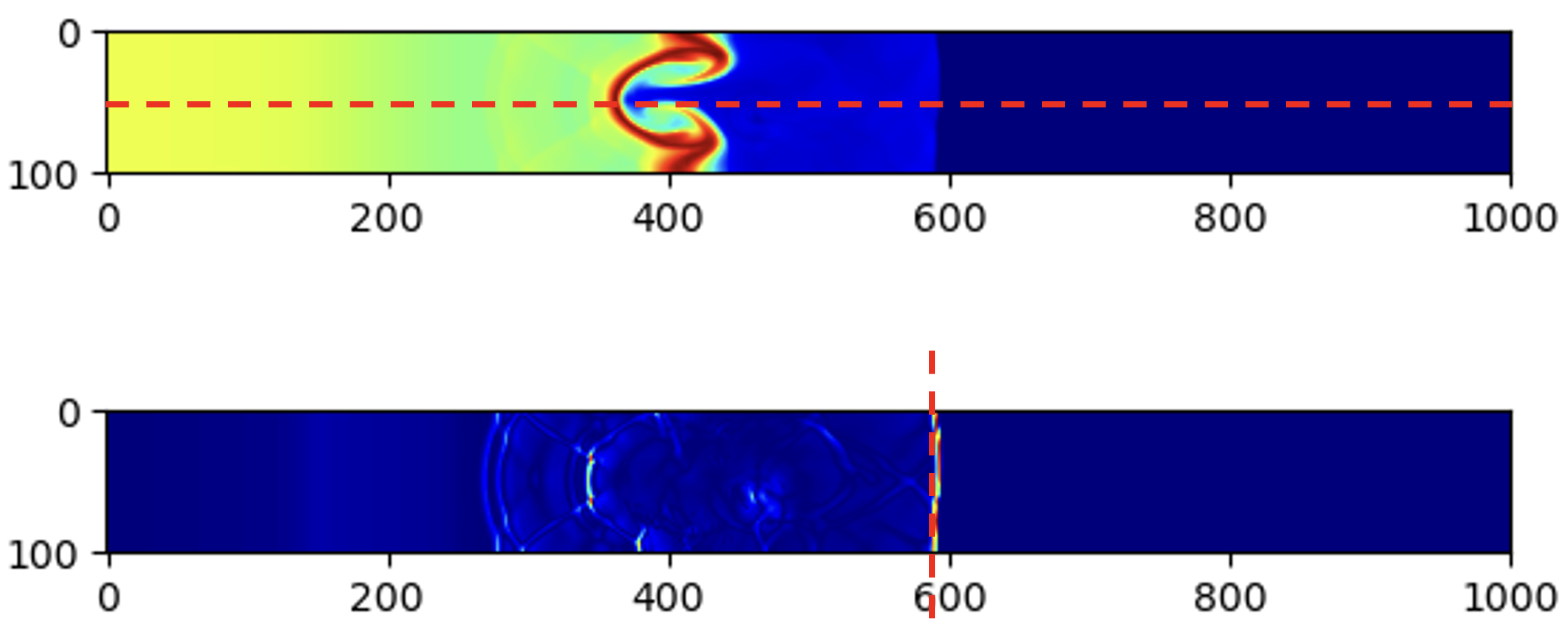}
    \caption{Illustration of quantities of interest studied in this work. Top: instantaneous temperature field. Quantities of interest $Q_{T_i}$ are the average temperatures along the centerline (red dashed line) at time step $i$. Bottom: instantaneous pressure gradient along $x$ direction (horizontal direction) field. The vertical red dash line highlights the shock front. Quantity of interest $Q_{p_0}$ is the maximum $x$-directional pressure gradient on the shock front. Quantity of interest $Q_{p_1}$ is the maximum $y$-directional pressure gradient on the shock front. }
    \label{fig: QOIs illustration}
\end{figure}

After finishing the simulations, quantities of interest are selected and extracted. In this section, we focus on two different QOIs, which are important indicators for the detonation structure in the channel. The first QOI is the average temperature along the centerline, which reflects how much heat is released and how well the fuel burns. Three different time instances are chosen, and they are 14, 28, and 40, respectively. Time step 14 is the moment short after the shock front passes the second droplet, and at this point, both droplets are detonated. At time 28, the detonation front reaches the end of the channel, and the temperature field is well-developed, while at time 40, the whole domain reaches a relatively high temperature. These three instances are chosen to reflect different stages of the detonation. Another important characteristic of the detonation structure is associated with the shock front. Maintaining a high strength of the shock front is of vital importance for detonation and engine performance. Thus, the second quantity of interest is chosen to be the maximum pressure gradient at the shock front. Since the pressure gradient is a vector, its two components are both considered. Fig.\ref{fig: QOIs illustration} provides an illustration of how these QOIs are extracted from the simulation field. 

\subsection{Single quantity of interest}

PCEs are built for all quantities of interest in the approach described in Figure~\ref{fig: PCE joint PDF framework}. The four random input variables are first converted from Beta distributions into standard Gaussian distributions using inverse CDF approach. Then, each QOI is expanded in the Gaussian germs. Since coefficients of polynomials are determined through least square regression and available samples are limited, it is important to prevent overfitting. Out of the total 1000 numerically simulated samples, we reserve 100 for testing, and construct the PCE model with the remaining 900 samples. Polynomial orders are increased until an obvious decrease in testing accuracy is detected, indicating the maximum polynomial order we can achieve with the current data. 

\begin{figure}[htb]
    \centering
    \begin{minipage}{0.45\textwidth}
        \centering
        \includegraphics[width=0.8\linewidth]{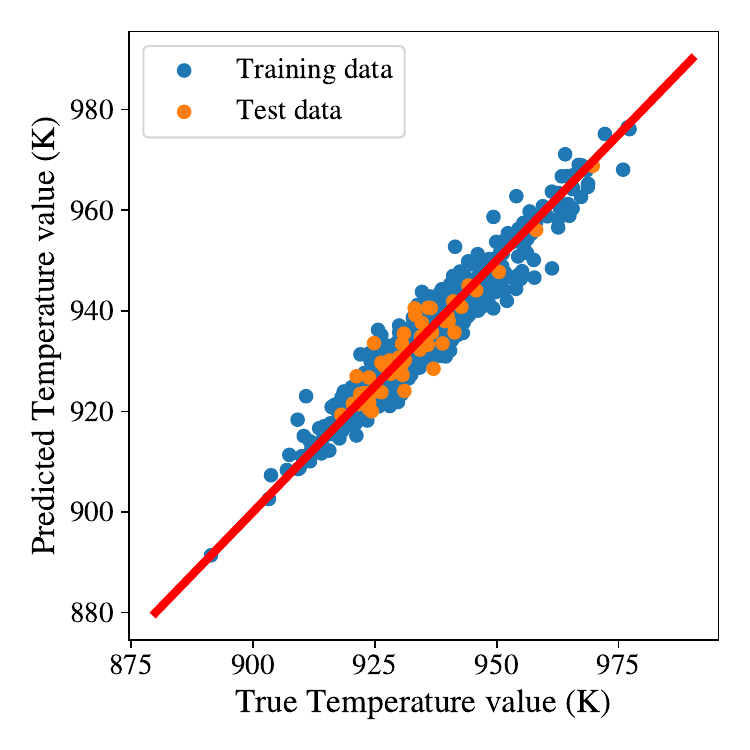}
    \end{minipage}
    \begin{minipage}{0.45\textwidth}
        \centering
        \includegraphics[width=0.8\linewidth]{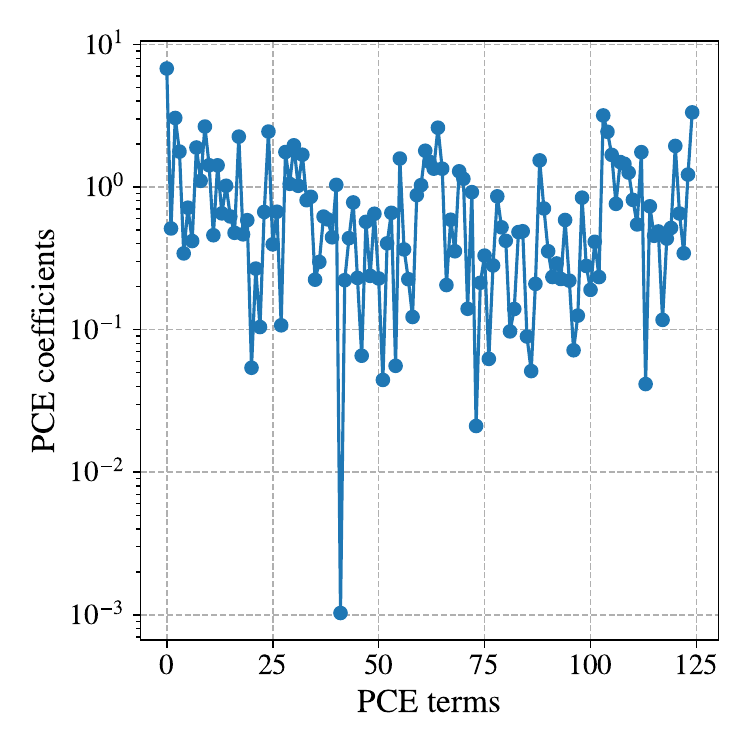}
    \end{minipage}
    \caption{PDF (top row) and CDF (bottom row) of maximum pressure gradient along the shock front at time step 15. Left panel: log-scale pressure gradient in $x$-direction. Right panel: log-scale pressure gradient in $y$-direction.}
    \label{fig: T average PCE}
\end{figure}

Figure~\ref{fig: T average PCE} shows the trained PCE for $Q_{T_{14}}$, i.e., the average temperature along the centerline at time instance 14. Fifth-order Hermite polynomials are chosen using the procedures described above. The left figure plots the comparison between predictions by PCE and ground truth for both training data and test data. The red diagonal line indicates the perfect prediction. It is obvious that the constructed PCE is a good surrogate model with accurate point-wise predictions. The right figure shows absolute values of PCE coefficients (excluding the constant term) in the log scale. Fifth-order full polynomial expansion in four-dimensional input space results in 126 terms in total. The first four points are coefficients corresponding to linear terms, followed by second-order terms, etc. One core observation is that higher-order terms are important in this expansion, as large values of coefficients are found for high-order polynomials. This is consistent with the well-known fact that turbulence is highly-nonlinear and sensitive to initial and boundary conditions, which typically requires high-order chaos expansion.  

\begin{figure}[htb]
    \centering
    \begin{minipage}{0.45\textwidth}
        \centering
        \includegraphics[width=0.7\linewidth]{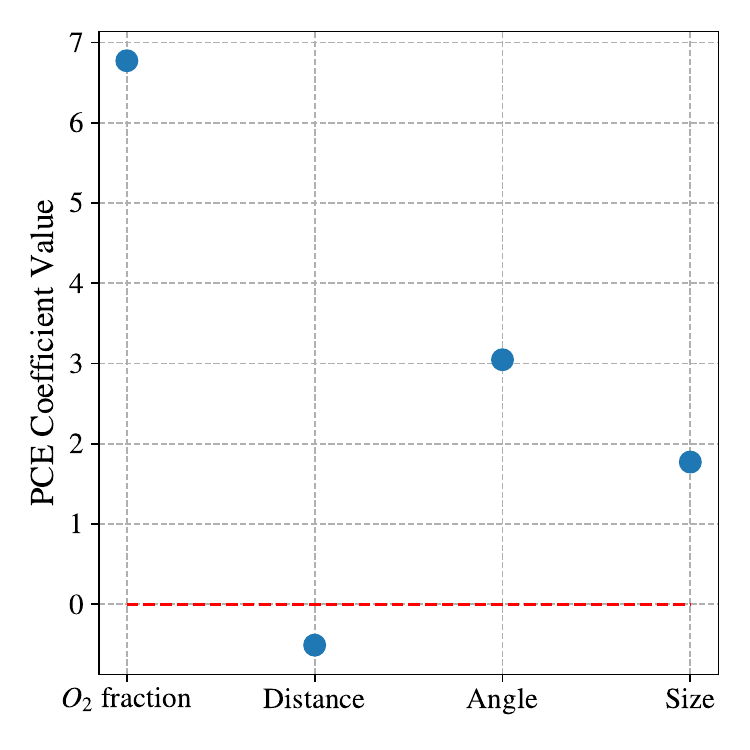}
    \end{minipage}
    \begin{minipage}{0.45\textwidth}
        \centering
        \includegraphics[width=0.8\linewidth]{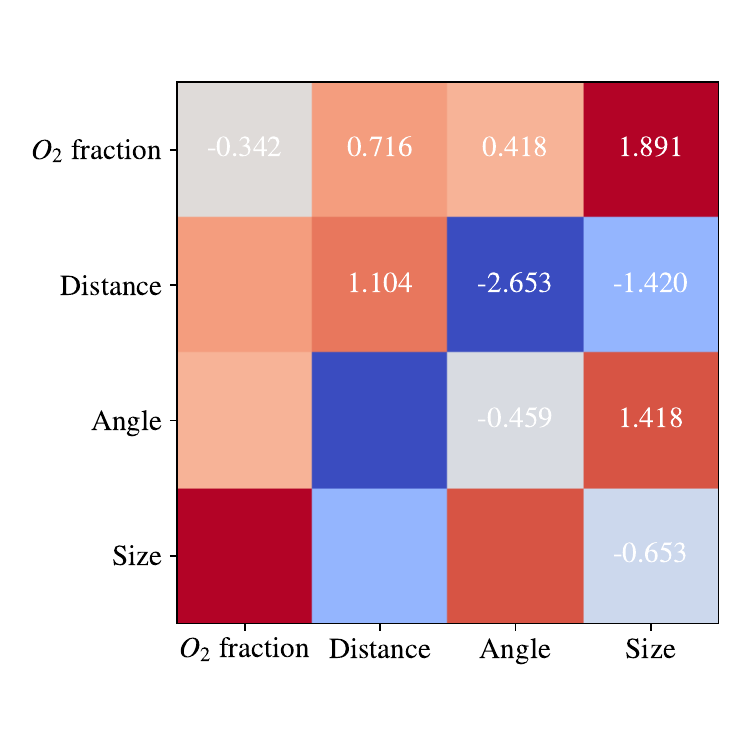}
    \end{minipage}
    \caption{First-order (left) and second-order (right) PCE coefficients. For second-order coefficients, large value is represented by red color while small value is represented by blue color.}
    \label{fig: PCE coef}
\end{figure}

To better understand the quantitative influence of input parameters on the quantity of interest, we examine the first- and second-order coefficients PCE terms, as shown in Figure~\ref{fig: PCE coef}. The linear (first-order) coefficients reveal that the oxygen volume fraction is the most dominant factor affecting $Q_{T_{14}}$. This result is consistent with physical intuition: under fuel-rich conditions, the extent of combustion—and consequently, the thermal energy released—is primarily limited by the availability of oxygen.
Among the four input parameters, all except the droplet distance exhibit positive linear correlations with $Q_{T_{14}}$. The second-order coefficients are visualized in a $4 \times 4$ matrix, where each entry corresponds to the interaction between a pair of input variables. Notably, the interaction between droplet spacing ($\Delta x$) and alignment angle ($\theta$) exhibits a large negative coefficient. This interaction term can be interpreted through the geometric approximation:
\begin{equation*}
    \Delta x \cdot \theta \approx \Delta x \cdot \tan(\theta) = \Delta y,
\end{equation*}
where $\Delta y$ represents the effective vertical misalignment between the two droplets. A smaller misalignment distance $\Delta y$ promotes stronger coupling between the shock wave and the second droplet, which facilitates the maintenance of a coherent detonation front. Conversely, increasing $\Delta y$ (through either greater spacing or larger angular misalignment) disrupts the symmetry and weakens detonation propagation.
It is crucial to note that relying solely on first-order sensitivity metrics, such as Sobol indices, may lead to misleading conclusions. For instance, the first-order coefficient of the alignment angle is positive, suggesting that increasing $\theta$ enhances $Q_{T_{14}}$. However, this contradicts both the second-order interaction analysis and our simulation results, which show that vertical misalignment generally deteriorates detonation strength. This discrepancy underscores the importance of incorporating higher-order terms in the surrogate model to fully capture nonlinear interactions and dependencies.

Overall, this analysis highlights the necessity of using high-order PCE for accurate uncertainty quantification in systems characterized by complex, multi-parameter interactions—such as those observed in droplet-detonation coupling.

\begin{figure}[htb]
    \centering
    \begin{minipage}{0.3\textwidth}
        \centering
        \includegraphics[width=\linewidth]{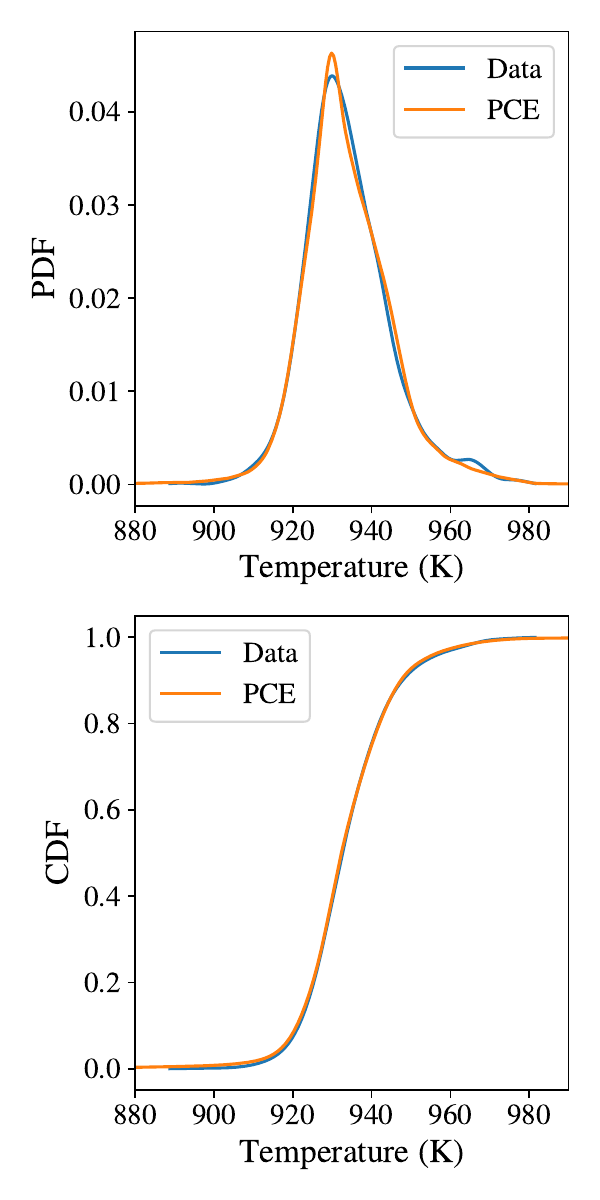}
    \end{minipage}
    \begin{minipage}{0.3\textwidth}
        \centering
        \includegraphics[width=\linewidth]{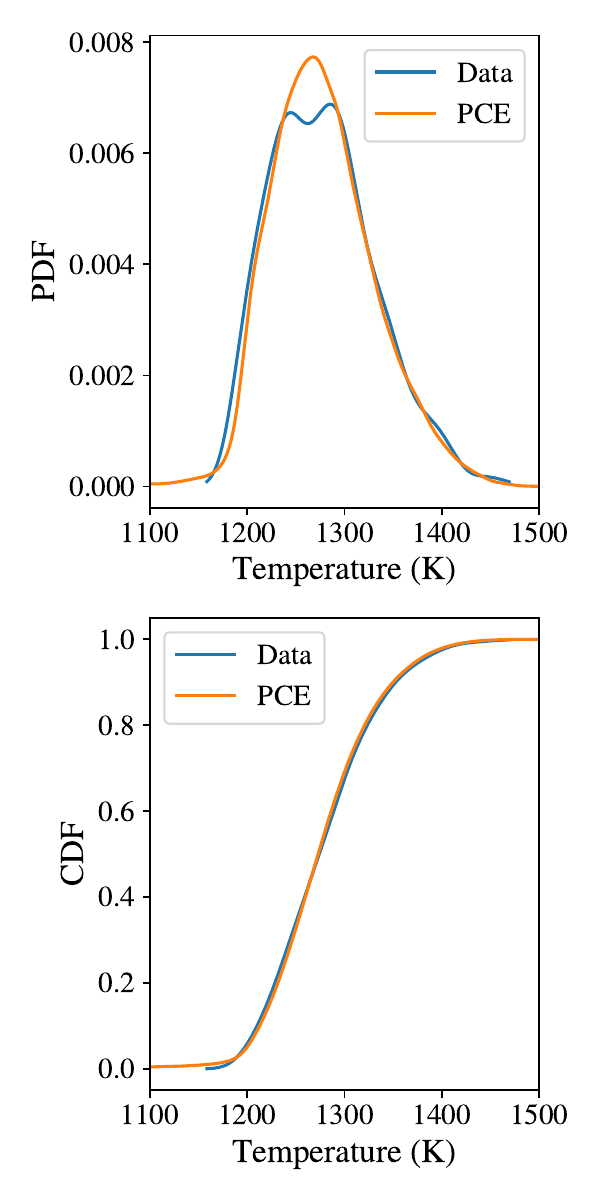}
    \end{minipage}
    \begin{minipage}{0.3\textwidth}
        \centering
        \includegraphics[width=\linewidth]{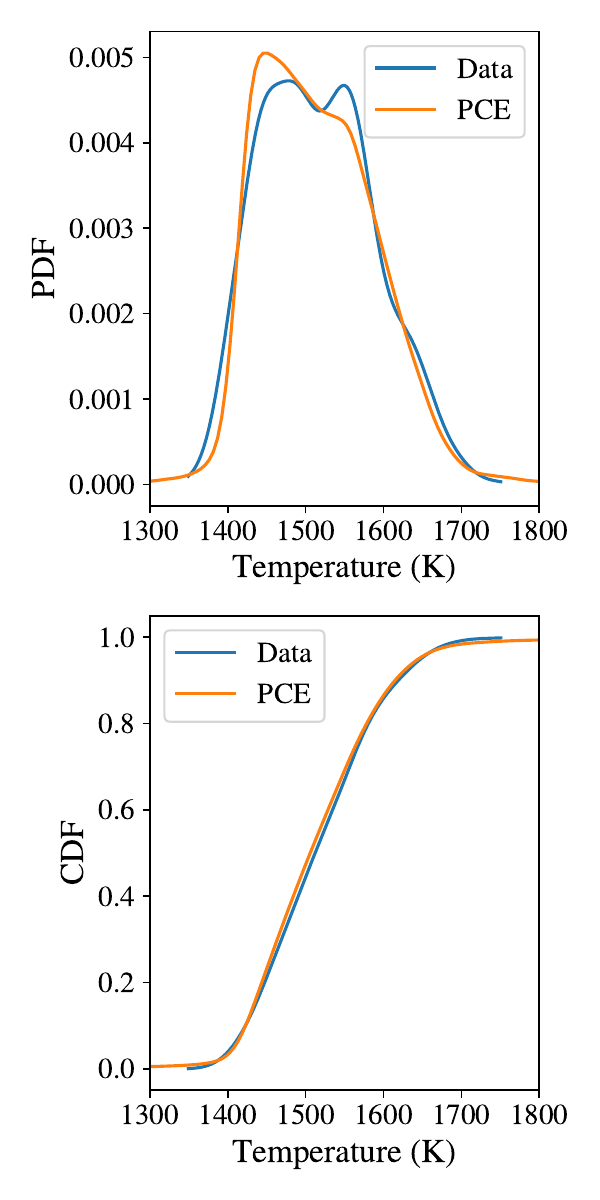}
    \end{minipage}
    \caption{PDF (top row) and CDF (bottom row) of average temperatures along the centerline of the computational domain at time step 14 (left panel), 28 (middle panel), and 40 (right panel).}
    \label{fig: pdf and cdf of QOIs temp}
\end{figure}

The PC expansions for $Q_{T_{28}}$ and $Q_{T_{40}}$, which represent the centerline-averaged temperature at time instances $t = 28$ and $t = 40$, respectively, are constructed following the same methodology as for $Q_{T_{14}}$. A fourth-order expansion is used for $Q_{T_{28}}$, while a fifth-order expansion is employed for $Q_{T_{40}}$. The probability density functions (PDFs) and cumulative distribution functions (CDFs) of all three quantities of interest ($Q_{T_{14}}$, $Q_{T_{28}}$, and $Q_{T_{40}}$) are presented in Figure~\ref{fig: pdf and cdf of QOIs temp}. The blue curves represent non-parametric estimates based on 1000 simulation samples using kernel density estimation (KDE), serving as a ground-truth reference. The orange curves correspond to the distributions estimated from the PCE surrogates, obtained by evaluating 40,000 independent samples drawn from the underlying standard Gaussian germs $\boldsymbol{\xi}$.
Overall, there is excellent agreement between the surrogate-based and simulation-based distributions, demonstrating that the PCE models are able to capture the dominant probabilistic structure of the system. However, for $Q_{T_{28}}$ and $Q_{T_{40}}$, the PDFs exhibit mild bi-modal characteristics near their peaks, which are only partially captured by the PCE approximations. This is a known limitation of polynomial expansions: accurately capturing multi-modality typically requires substantially higher-order terms, which is constrained here due to the limited number of available high-fidelity simulations.
Moreover, a gradual degradation in surrogate performance is observed as time progresses—from $t = 14$ to $t = 40$—indicating increased complexity and variability in the temperature field during the evolution of the detonation. This trend highlights the growing importance of higher-order nonlinear interactions in governing the quantities of interest over time. Nevertheless, the PCE surrogates successfully reproduce the core features of the distributions, particularly for early- to mid-stage detonation behavior. These results reinforce the capability of polynomial chaos methods as efficient and interpretable tools for uncertainty quantification in transient, high-speed reactive flow systems.

\begin{figure}[htb]
    \centering
    \begin{minipage}{0.45\textwidth}
        \centering
        \includegraphics[width=0.8\linewidth]{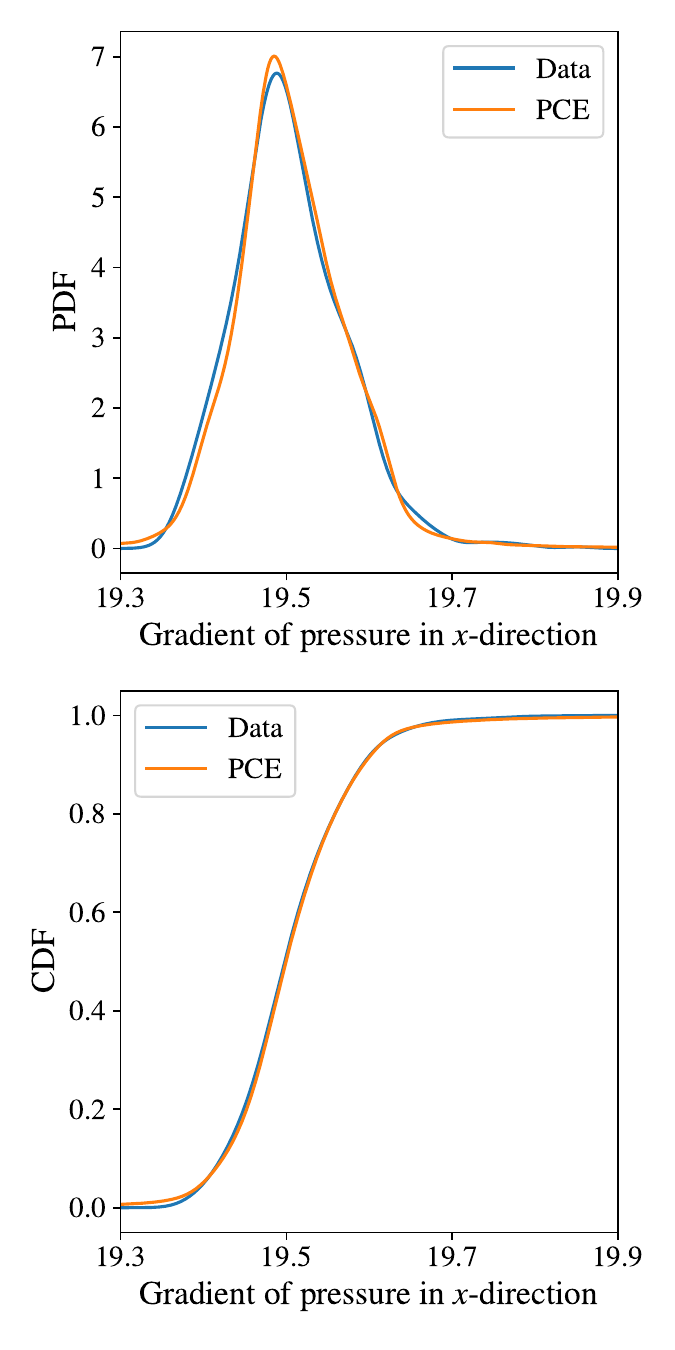}
    \end{minipage}
    \begin{minipage}{0.45\textwidth}
        \centering
        \includegraphics[width=0.8\linewidth]{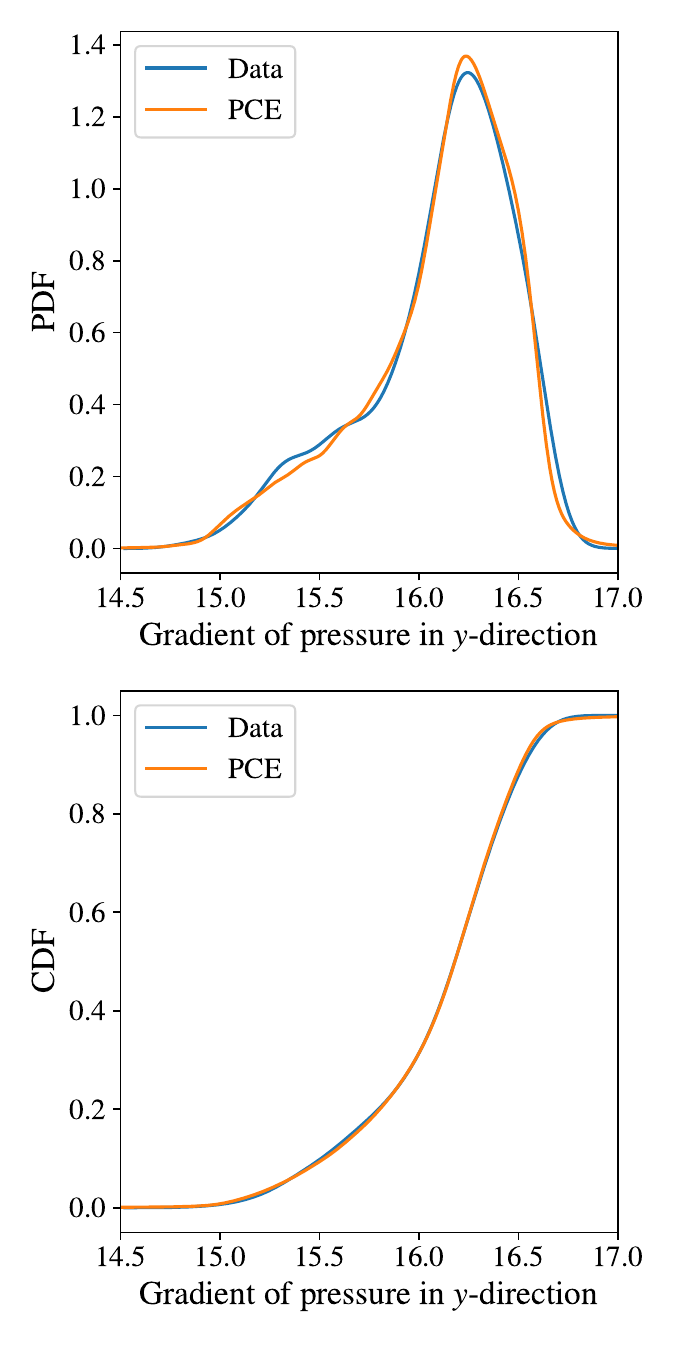}
    \end{minipage}
    \caption{PDF (top row) and CDF (bottom row) of maximum pressure gradient along the shock front at time step 15. Left panel: log-scale pressure gradient in $x$-direction. Right panel: log-scale pressure gradient in $y$-direction}
    \label{fig: grad p 0 and grad p1 pdf and cdf}
\end{figure}

While temperature information gives us one important perspective of the detonation process, the shock front plays a more significant role in the characterization of detonations. The two quantities of interest $Q_{p_{0}}$ and $Q_{p_{1}}$, reflecting the strength of the shock front, are extracted, and their PCEs are constructed. Since the pressure gradient values are typically large, with $10^6$ magnitude, the log values are taken instead of their original values. Figure~\ref{fig: grad p 0 and grad p1 pdf and cdf} shows the PDFs and CDFs from PCEs and from simulation data. Both QOIs are captured well by fifth-order PCEs. 

\subsubsection{Joint quantities of interest}
\begin{figure}[htb]
    \centering
    \includegraphics[width=0.9\linewidth]{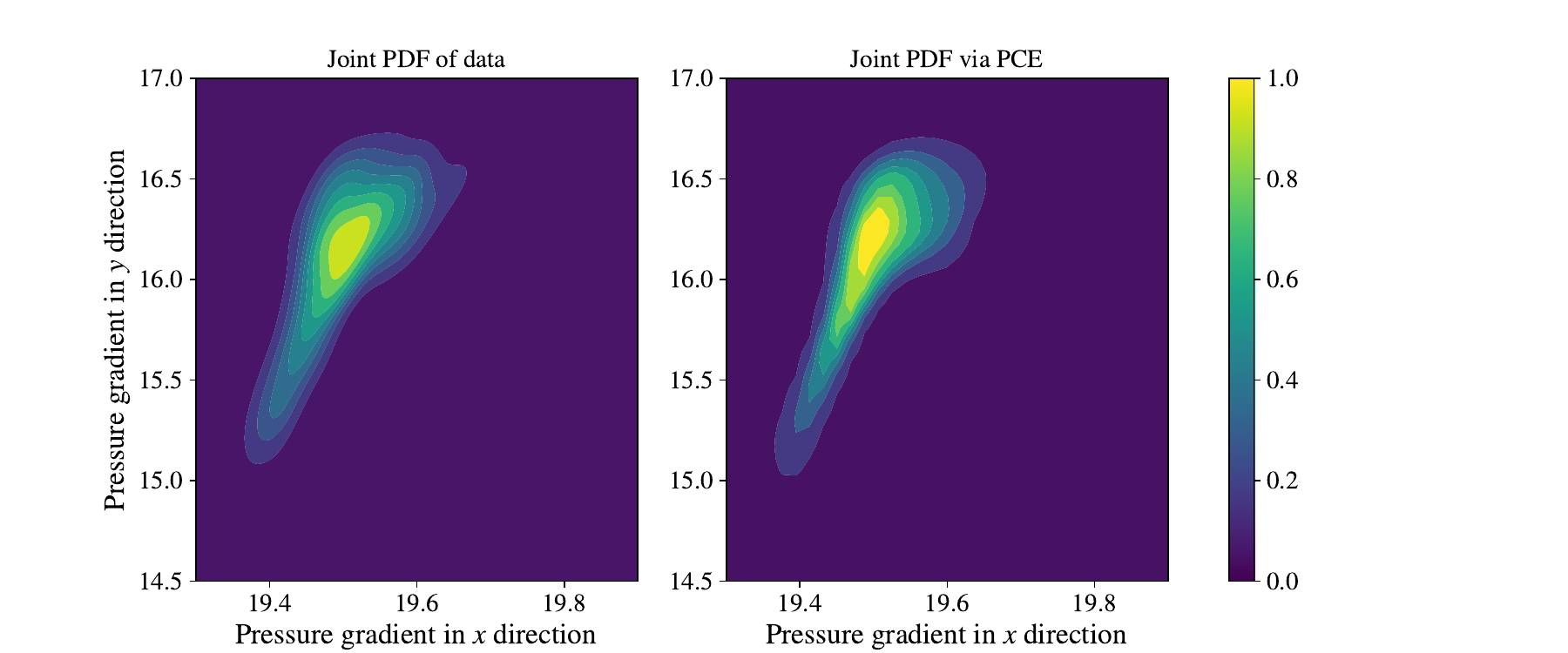}
    \caption{Joint distribution of maximum pressure gradient on the shock front in $x$ and $y$ direction. Left: numerically simulated data from physical code; right: PCE-generated data.}
    \label{fig: joint pdf of grad p0 and p1}
\end{figure}

The previous section builds the PCE for each quantity of interest on the same chaos space, which provides a unified basis for the characterization of joint distributions of QOIs. As depicted in Figure~\ref{fig: PCE joint PDF framework}, we sample 40000 independent Gaussian germs from $\boldsymbol{\xi}$, then evaluate all the PCEs at sampled points, resulting in 40000 joint occurrences of QOIs, from which joint distributions can be characterized in usual ways. The joint distribution between $Q_{p_{0}}$ and $Q_{p_{1}}$ is plotted in Figure~\ref{fig: joint pdf of grad p0 and p1}. The left plot comes from 1000 simulation data and the right one from 40000 PCE samples. It is clear that our proposed framework yields accurate joint distributions. In addition, the joint distributions show that gradients in the $x$ direction and $y$ direction have a strong positive correlation when the magnitude is small, while this correlation gets weaker when the gradient value is large. Similar procedures can be applied to other pairwise joint distributions as well. What's more, this approach makes it possible to study joint distributions of three or more random variables, which is difficult to quantify using only existing simulation data. 

\section{Discussion and Conclusion}

The detonation simulations investigated in this work share similarities but also exhibit fundamental differences compared to existing studies on gaseous detonation. Previous studies, such as those by \citet{yan2021effects, crane2023three, paknahad2024statistics, chen2024implementation}, have provided detailed examinations of detonation structures within fully premixed, homogeneous gaseous mixtures filling the entire combustion chamber. In these idealized scenarios, the characteristic cellular structures, commonly described as "fishnet" patterns, have been clearly observed and quantitatively analyzed. In contrast, practical detonation engines typically operate under conditions where fuel distribution is inherently non-homogeneous due to fuel injection techniques, incomplete mixing, and geometric constraints within the combustion chamber \citep{raj2024influence}. Such inhomogeneity significantly influences the propagation and stability of detonation waves, differentiating realistic engine conditions from idealized homogeneous assumptions. In our current work, we have adopted a simplified yet physically insightful approach, embedding discrete gas-phase fuel droplets composed of premixed hydrogen–oxygen into a computational domain initially filled with ambient air. A gaseous detonation wave is initiated within an ignition zone and propagates toward the droplets, subsequently detonating them and thereby sustaining and intensifying the propagation of the detonation front. This controlled setting enables isolation and rigorous examination of shock–droplet interactions within the gas phase, significantly advancing our understanding of more complicated scenarios relevant to practical detonation engines. Despite the apparent simplicity, the observed flow structures exhibit considerable complexity. We have identified unique and intricate patterns illustrating how the detonation of one droplet alters flow dynamics and how subsequent droplets contribute in a nonlinear and cumulative manner. Furthermore, our results underscore the importance of droplet spacing and alignment, providing direct insights applicable to droplet injection strategies within actual engine designs. The capability of extending our simulation framework to more droplets also suggests broad applicability for future parametric studies.

It is also noteworthy to distinguish our gas-phase droplet detonation scenario from liquid droplet detonation, an area of intensive contemporary research. Liquid droplet detonation involves additional complexities, including droplet evaporation \citep{wen2023numerical}, shock–liquid interactions \citep{prakash2024three, xu2024numerical}, droplet deformation, breakup, and secondary atomization \citep{srinivasan2024vle, xu2024numerical, dammati2025numerical}. Such phenomena necessitate computationally demanding multiphase simulations. In contrast, by assuming gas-phase droplets, our investigation eliminates these complexities and focuses explicitly on gas-phase detonative interactions, thus enabling detailed and computationally efficient analyses of core detonation physics. 

In summary, the presence of fuel droplets in our simulations significantly intensifies the forward-moving detonation front. Moreover, it induces a notable backward-propagating detonation wave, whose strength is further enhanced by each additional fuel droplet. We found this phenomenon to be sensitive to droplet characteristics such as inter-droplet distance, horizontal and vertical alignment, fuel composition, and droplet size. Quantitative uncertainty analysis via PCE further revealed that capturing these intricate physics requires modeling higher-order interactions between parameters. The importance of incorporating such higher-order dependencies underscores the complexity inherent to detonation phenomena.
As this study contributes to the exploration of rotating detonation engines (RDEs) featuring discrete fuel injections, it highlights critical factors influencing engine performance, particularly the geometry and configuration of fuel inlets. Through detailed characterization of these parameters, we reveal the intricate physics governing detonation wave propagation, the complexities of real-world engine design, and the inherent uncertainties that must be managed. Whether powering next-generation aircraft, spacecraft, or industrial machinery, the RDE with optimized fuel inlet geometry presents a compelling vision for the future of propulsion technology.


\bibliographystyle{elsarticle-harv} 
\bibliography{References.bib}

\begin{thebibliography}{41}
\expandafter\ifx\csname natexlab\endcsname\relax\def\natexlab#1{#1}\fi
\providecommand{\url}[1]{\texttt{#1}}
\providecommand{\href}[2]{#2}
\providecommand{\path}[1]{#1}
\providecommand{\DOIprefix}{doi:}
\providecommand{\ArXivprefix}{arXiv:}
\providecommand{\URLprefix}{URL: }
\providecommand{\Pubmedprefix}{pmid:}
\providecommand{\doi}[1]{\href{http://dx.doi.org/#1}{\path{#1}}}
\providecommand{\Pubmed}[1]{\href{pmid:#1}{\path{#1}}}
\providecommand{\bibinfo}[2]{#2}
\ifx\xfnm\relax \def\xfnm[#1]{\unskip,\space#1}\fi
\bibitem[{Back(1975)}]{back1975application}
\bibinfo{author}{Back, L.H.}, \bibinfo{year}{1975}.
\newblock \bibinfo{title}{Application of blast wave theory to explosive propulsion}.
\newblock \bibinfo{journal}{Acta Astronautica} \bibinfo{volume}{2}, \bibinfo{pages}{391--407}.
\bibitem[{Betelin et~al.(2020)Betelin, Nikitin and Mikhalchenko}]{betelin20203d}
\bibinfo{author}{Betelin, V.B.}, \bibinfo{author}{Nikitin, V.F.}, \bibinfo{author}{Mikhalchenko, E.V.}, \bibinfo{year}{2020}.
\newblock \bibinfo{title}{3d numerical modeling of a cylindrical rde with an inner body extending out of the nozzle}.
\newblock \bibinfo{journal}{Acta Astronautica} \bibinfo{volume}{176}, \bibinfo{pages}{628--646}.
\bibitem[{Bluemner et~al.(2020)Bluemner, Bohon, Paschereit and Gutmark}]{bluemner2020effect}
\bibinfo{author}{Bluemner, R.}, \bibinfo{author}{Bohon, M.D.}, \bibinfo{author}{Paschereit, C.O.}, \bibinfo{author}{Gutmark, E.J.}, \bibinfo{year}{2020}.
\newblock \bibinfo{title}{Effect of inlet and outlet boundary conditions on rotating detonation combustion}.
\newblock \bibinfo{journal}{Combustion and Flame} \bibinfo{volume}{216}, \bibinfo{pages}{300--315}.
\bibitem[{Chen et~al.(2024)Chen, Zhao, Qiu and Zhu}]{chen2024implementation}
\bibinfo{author}{Chen, H.}, \bibinfo{author}{Zhao, M.}, \bibinfo{author}{Qiu, H.}, \bibinfo{author}{Zhu, Y.}, \bibinfo{year}{2024}.
\newblock \bibinfo{title}{Implementation and verification of an openfoam solver for gas-droplet two-phase detonation combustion}.
\newblock \bibinfo{journal}{Physics of Fluids} \bibinfo{volume}{36}.
\bibitem[{Crane et~al.(2023)Crane, Lipkowicz, Shi, Wlokas, Kempf and Wang}]{crane2023three}
\bibinfo{author}{Crane, J.}, \bibinfo{author}{Lipkowicz, J.T.}, \bibinfo{author}{Shi, X.}, \bibinfo{author}{Wlokas, I.}, \bibinfo{author}{Kempf, A.M.}, \bibinfo{author}{Wang, H.}, \bibinfo{year}{2023}.
\newblock \bibinfo{title}{Three-dimensional detonation structure and its response to confinement}.
\newblock \bibinfo{journal}{Proceedings of the combustion institute} \bibinfo{volume}{39}, \bibinfo{pages}{2915--2923}.
\bibitem[{Dammati et~al.(2025)Dammati, Poludnenko, Kateris, Dong, Wang and Lu}]{dammati2025numerical}
\bibinfo{author}{Dammati, S.S.}, \bibinfo{author}{Poludnenko, A.}, \bibinfo{author}{Kateris, N.}, \bibinfo{author}{Dong, W.}, \bibinfo{author}{Wang, H.}, \bibinfo{author}{Lu, T.}, \bibinfo{year}{2025}.
\newblock \bibinfo{title}{Numerical simulations of non-ideal spray detonations in jet fuels with a shock-droplet interaction model}, in: \bibinfo{booktitle}{AIAA SCITECH 2025 Forum}, p. \bibinfo{pages}{0388}.
\bibitem[{Ghanem(1989)}]{spanos1989stochastic}
\bibinfo{author}{Ghanem, R.}, \bibinfo{year}{1989}.
\newblock \bibinfo{title}{Stochastic finite element expansion for random media}.
\newblock \bibinfo{journal}{Journal of engineering mechanics} \bibinfo{volume}{115}, \bibinfo{pages}{1035--1053}.
\bibitem[{Ghiocel and Ghanem(2002)}]{ghiocel2002stochastic}
\bibinfo{author}{Ghiocel, D.M.}, \bibinfo{author}{Ghanem, R.G.}, \bibinfo{year}{2002}.
\newblock \bibinfo{title}{Stochastic finite-element analysis of seismic soil--structure interaction}.
\newblock \bibinfo{journal}{Journal of Engineering Mechanics} \bibinfo{volume}{128}, \bibinfo{pages}{66--77}.
\bibitem[{Gordon et~al.(1958)Gordon, Mooradian and Harper}]{gordon1958limit}
\bibinfo{author}{Gordon, W.}, \bibinfo{author}{Mooradian, A.}, \bibinfo{author}{Harper, S.}, \bibinfo{year}{1958}.
\newblock \bibinfo{title}{Limit and spin effects in hydrogen-oxygen detonations}, in: \bibinfo{booktitle}{Symposium (International) on Combustion}, \bibinfo{organization}{Elsevier}. pp. \bibinfo{pages}{752--759}.
\bibitem[{Greenshields(2023)}]{greenshields2023}
\bibinfo{author}{Greenshields, C.}, \bibinfo{year}{2023}.
\newblock \bibinfo{title}{OpenFOAM v11 User Guide}.
\newblock \bibinfo{publisher}{The OpenFOAM Foundation}, \bibinfo{address}{London, UK}.
\newblock \URLprefix \url{https://doc.cfd.direct/openfoam/user-guide-v11}.
\bibitem[{Hadigol and Doostan(2018)}]{hadigol2018least}
\bibinfo{author}{Hadigol, M.}, \bibinfo{author}{Doostan, A.}, \bibinfo{year}{2018}.
\newblock \bibinfo{title}{Least squares polynomial chaos expansion: A review of sampling strategies}.
\newblock \bibinfo{journal}{Computer Methods in Applied Mechanics and Engineering} \bibinfo{volume}{332}, \bibinfo{pages}{382--407}.
\bibitem[{Hampton and Doostan(2015)}]{hampton2015coherence}
\bibinfo{author}{Hampton, J.}, \bibinfo{author}{Doostan, A.}, \bibinfo{year}{2015}.
\newblock \bibinfo{title}{Coherence motivated sampling and convergence analysis of least squares polynomial chaos regression}.
\newblock \bibinfo{journal}{Computer Methods in Applied Mechanics and Engineering} \bibinfo{volume}{290}, \bibinfo{pages}{73--97}.
\bibitem[{Humphrey(1909)}]{humphrey1909internal}
\bibinfo{author}{Humphrey, H.A.}, \bibinfo{year}{1909}.
\newblock \bibinfo{title}{An internal-combustion pump, and other applications of a new principle}.
\newblock \bibinfo{journal}{Proceedings of the Institution of Mechanical Engineers} \bibinfo{volume}{77}, \bibinfo{pages}{1075--1200}.
\bibitem[{Kailasanath(2011)}]{kailasanath2011rotating}
\bibinfo{author}{Kailasanath, K.}, \bibinfo{year}{2011}.
\newblock \bibinfo{title}{The rotating detonation-wave engine concept: A brief status report}, in: \bibinfo{booktitle}{49th AIAA aerospace sciences meeting including the new horizons forum and aerospace exposition}, p. \bibinfo{pages}{580}.
\bibitem[{Kindracki et~al.(2020)Kindracki, Siatkowski and Lukasik}]{kindracki2020influence}
\bibinfo{author}{Kindracki, J.}, \bibinfo{author}{Siatkowski, S.}, \bibinfo{author}{Lukasik, B.}, \bibinfo{year}{2020}.
\newblock \bibinfo{title}{Influence of inlet flow parameters on rotating detonation}.
\newblock \bibinfo{journal}{AIAA Journal} \bibinfo{volume}{58}, \bibinfo{pages}{5046--5051}.
\bibitem[{Knio et~al.(2001)Knio, Najm, Ghanem et~al.}]{knio2001stochastic}
\bibinfo{author}{Knio, O.M.}, \bibinfo{author}{Najm, H.N.}, \bibinfo{author}{Ghanem, R.G.}, et~al., \bibinfo{year}{2001}.
\newblock \bibinfo{title}{A stochastic projection method for fluid flow: I. basic formulation}.
\newblock \bibinfo{journal}{Journal of computational Physics} \bibinfo{volume}{173}, \bibinfo{pages}{481--511}.
\bibitem[{Le~Ma{\i}tre et~al.(2002)Le~Ma{\i}tre, Reagan, Najm, Ghanem and Knio}]{le2002stochastic}
\bibinfo{author}{Le~Ma{\i}tre, O.P.}, \bibinfo{author}{Reagan, M.T.}, \bibinfo{author}{Najm, H.N.}, \bibinfo{author}{Ghanem, R.G.}, \bibinfo{author}{Knio, O.M.}, \bibinfo{year}{2002}.
\newblock \bibinfo{title}{A stochastic projection method for fluid flow: Ii. random process}.
\newblock \bibinfo{journal}{Journal of computational Physics} \bibinfo{volume}{181}, \bibinfo{pages}{9--44}.
\bibitem[{Lu and Braun(2014)}]{lu2014rotating}
\bibinfo{author}{Lu, F.K.}, \bibinfo{author}{Braun, E.M.}, \bibinfo{year}{2014}.
\newblock \bibinfo{title}{Rotating detonation wave propulsion: experimental challenges, modeling, and engine concepts}.
\newblock \bibinfo{journal}{Journal of Propulsion and Power} \bibinfo{volume}{30}, \bibinfo{pages}{1125--1142}.
\bibitem[{Ma et~al.(2018)Ma, Zhang, Luan, Yao, Xia and Wang}]{ma2018experimental}
\bibinfo{author}{Ma, Z.}, \bibinfo{author}{Zhang, S.}, \bibinfo{author}{Luan, M.}, \bibinfo{author}{Yao, S.}, \bibinfo{author}{Xia, Z.}, \bibinfo{author}{Wang, J.}, \bibinfo{year}{2018}.
\newblock \bibinfo{title}{Experimental research on ignition, quenching, reinitiation and the stabilization process in rotating detonation engine}.
\newblock \bibinfo{journal}{International journal of hydrogen energy} \bibinfo{volume}{43}, \bibinfo{pages}{18521--18529}.
\bibitem[{Marcantoni et~al.(2017)Marcantoni, Tamagno and Elaskar}]{marcantoni2017rhocentralrffoam}
\bibinfo{author}{Marcantoni, L.F.G.}, \bibinfo{author}{Tamagno, J.}, \bibinfo{author}{Elaskar, S.}, \bibinfo{year}{2017}.
\newblock \bibinfo{title}{{rhoCentralRfFoam:} an openfoam solver for high speed chemically active flows--simulation of planar detonations--}.
\newblock \bibinfo{journal}{Computer Physics Communications} \bibinfo{volume}{219}, \bibinfo{pages}{209--222}.
\bibitem[{Marinov et~al.(1996)Marinov, Westbrook and Pitz}]{marinov1996detailed}
\bibinfo{author}{Marinov, N.M.}, \bibinfo{author}{Westbrook, C.K.}, \bibinfo{author}{Pitz, W.J.}, \bibinfo{year}{1996}.
\newblock \bibinfo{title}{Detailed and global chemical kinetics model for hydrogen}.
\newblock \bibinfo{journal}{Transport phenomena in combustion} \bibinfo{volume}{1}, \bibinfo{pages}{80}.
\bibitem[{Nicholls et~al.(1957)Nicholls, Wilkinson and Morrison}]{nicholls1957intermittent}
\bibinfo{author}{Nicholls, J.A.}, \bibinfo{author}{Wilkinson, H.R.}, \bibinfo{author}{Morrison, R.B.}, \bibinfo{year}{1957}.
\newblock \bibinfo{title}{Intermittent detonation as a thrust-producing mechanism}.
\newblock \bibinfo{journal}{Journal of jet propulsion} \bibinfo{volume}{27}, \bibinfo{pages}{534--541}.
\bibitem[{Paknahad et~al.(2024)Paknahad, Lipkowicz, Gaffran, Wlokas, Kempf and Crane}]{paknahad2024statistics}
\bibinfo{author}{Paknahad, R.}, \bibinfo{author}{Lipkowicz, J.T.}, \bibinfo{author}{Gaffran, N.}, \bibinfo{author}{Wlokas, I.}, \bibinfo{author}{Kempf, A.M.}, \bibinfo{author}{Crane, J.}, \bibinfo{year}{2024}.
\newblock \bibinfo{title}{Statistics of detonation confinement: 1d, 2d and 3d simulations in hydrogen--oxygen}.
\newblock \bibinfo{journal}{Proceedings of the Combustion Institute} \bibinfo{volume}{40}, \bibinfo{pages}{105388}.
\bibitem[{Palaniswamy et~al.(2018)Palaniswamy, Akdag, Peroomian and Chakravarthy}]{palaniswamy2018comparison}
\bibinfo{author}{Palaniswamy, S.}, \bibinfo{author}{Akdag, V.}, \bibinfo{author}{Peroomian, O.}, \bibinfo{author}{Chakravarthy, S.}, \bibinfo{year}{2018}.
\newblock \bibinfo{title}{Comparison between ideal and slot injection in a rotating detonation engine}.
\newblock \bibinfo{journal}{Combustion Science and Technology} \bibinfo{volume}{190}, \bibinfo{pages}{557--578}.
\bibitem[{Prakash et~al.(2024)Prakash, Bielawski, Raman, Ahmed and Bennewitz}]{prakash2024three}
\bibinfo{author}{Prakash, S.}, \bibinfo{author}{Bielawski, R.}, \bibinfo{author}{Raman, V.}, \bibinfo{author}{Ahmed, K.}, \bibinfo{author}{Bennewitz, J.}, \bibinfo{year}{2024}.
\newblock \bibinfo{title}{Three-dimensional numerical simulations of a liquid rp-2/o2 based rotating detonation engine}.
\newblock \bibinfo{journal}{Combustion and Flame} \bibinfo{volume}{259}, \bibinfo{pages}{113097}.
\bibitem[{Raj and Meadows(2024)}]{raj2024influence}
\bibinfo{author}{Raj, P.}, \bibinfo{author}{Meadows, J.}, \bibinfo{year}{2024}.
\newblock \bibinfo{title}{Influence of fuel inhomogeneity on detonation wave propagation in a rotating detonation combustor}.
\newblock \bibinfo{journal}{Shock Waves} \bibinfo{volume}{34}, \bibinfo{pages}{429--449}.
\bibitem[{Raman et~al.(2023)Raman, Prakash and Gamba}]{raman2023nonidealities}
\bibinfo{author}{Raman, V.}, \bibinfo{author}{Prakash, S.}, \bibinfo{author}{Gamba, M.}, \bibinfo{year}{2023}.
\newblock \bibinfo{title}{Nonidealities in rotating detonation engines}.
\newblock \bibinfo{journal}{Annual Review of Fluid Mechanics} \bibinfo{volume}{55}, \bibinfo{pages}{639--674}.
\bibitem[{Rankin et~al.(2017)Rankin, Fotia, Naples, Stevens, Hoke, Kaemming, Theuerkauf and Schauer}]{rankin2017overview}
\bibinfo{author}{Rankin, B.A.}, \bibinfo{author}{Fotia, M.L.}, \bibinfo{author}{Naples, A.G.}, \bibinfo{author}{Stevens, C.A.}, \bibinfo{author}{Hoke, J.L.}, \bibinfo{author}{Kaemming, T.A.}, \bibinfo{author}{Theuerkauf, S.W.}, \bibinfo{author}{Schauer, F.R.}, \bibinfo{year}{2017}.
\newblock \bibinfo{title}{Overview of performance, application, and analysis of rotating detonation engine technologies}.
\newblock \bibinfo{journal}{Journal of Propulsion and Power} \bibinfo{volume}{33}, \bibinfo{pages}{131--143}.
\bibitem[{Reagan et~al.(2003)Reagan, Najm, Ghanem and Knio}]{reagana2003uncertainty}
\bibinfo{author}{Reagan, M.T.}, \bibinfo{author}{Najm, H.N.}, \bibinfo{author}{Ghanem, R.G.}, \bibinfo{author}{Knio, O.M.}, \bibinfo{year}{2003}.
\newblock \bibinfo{title}{Uncertainty quantification in reacting-flow simulations through non-intrusive spectral projection}.
\newblock \bibinfo{journal}{Combustion and flame} \bibinfo{volume}{132}, \bibinfo{pages}{545--555}.
\bibitem[{Roy(1946)}]{roy1946propulsion}
\bibinfo{author}{Roy, M.}, \bibinfo{year}{1946}.
\newblock \bibinfo{title}{Propulsion par statoreacteur a detonation}.
\newblock \bibinfo{journal}{Comptes Rendus Hebdomadaires des S{\'e}ances de l’Acad{\'e}mie des Sciences} \bibinfo{volume}{222}, \bibinfo{pages}{31--32}.
\bibitem[{Schwer and Kailasanath(2011)}]{schwer2011effect}
\bibinfo{author}{Schwer, D.}, \bibinfo{author}{Kailasanath, K.}, \bibinfo{year}{2011}.
\newblock \bibinfo{title}{Effect of inlet on fill region and performance of rotating detonation engines}, in: \bibinfo{booktitle}{47th AIAA/ASME/SAE/ASEE joint propulsion conference \& exhibit}, p. \bibinfo{pages}{6044}.
\bibitem[{Soize and Ghanem(2004)}]{soize2004physical}
\bibinfo{author}{Soize, C.}, \bibinfo{author}{Ghanem, R.}, \bibinfo{year}{2004}.
\newblock \bibinfo{title}{Physical systems with random uncertainties: chaos representations with arbitrary probability measure}.
\newblock \bibinfo{journal}{SIAM Journal on Scientific Computing} \bibinfo{volume}{26}, \bibinfo{pages}{395--410}.
\bibitem[{Srinivasan et~al.(2024)Srinivasan, Suryanarayan, Zhang, Ghosh, Dammati, Poludnenko and Yang}]{srinivasan2024vle}
\bibinfo{author}{Srinivasan, N.}, \bibinfo{author}{Suryanarayan, R.}, \bibinfo{author}{Zhang, H.}, \bibinfo{author}{Ghosh, A.}, \bibinfo{author}{Dammati, S.S.}, \bibinfo{author}{Poludnenko, A.}, \bibinfo{author}{Yang, S.}, \bibinfo{year}{2024}.
\newblock \bibinfo{title}{Vle-based high pressure stationary droplet evaporation at spray detonation conditions}, in: \bibinfo{booktitle}{AIAA SCITECH 2024 Forum}, p. \bibinfo{pages}{1636}.
\bibitem[{Sun et~al.(2017)Sun, Zhou, Liu, Lin and Cai}]{sun2017effects}
\bibinfo{author}{Sun, J.}, \bibinfo{author}{Zhou, J.}, \bibinfo{author}{Liu, S.}, \bibinfo{author}{Lin, Z.}, \bibinfo{author}{Cai, J.}, \bibinfo{year}{2017}.
\newblock \bibinfo{title}{Effects of injection nozzle exit width on rotating detonation engine}.
\newblock \bibinfo{journal}{Acta Astronautica} \bibinfo{volume}{140}, \bibinfo{pages}{388--401}.
\bibitem[{Tipireddy and Ghanem(2014)}]{Tipireddy:2014}
\bibinfo{author}{Tipireddy, R.}, \bibinfo{author}{Ghanem, R.}, \bibinfo{year}{2014}.
\newblock \bibinfo{title}{Basis adaptation in homogeneous chaos spaces}.
\newblock \bibinfo{journal}{Journal of Computational Physics} \bibinfo{volume}{259}, \bibinfo{pages}{304--317}.
\bibitem[{Varsi et~al.(1976)Varsi, Back and Kim}]{varsi1976blast}
\bibinfo{author}{Varsi, G.}, \bibinfo{author}{Back, L.H.}, \bibinfo{author}{Kim, K.}, \bibinfo{year}{1976}.
\newblock \bibinfo{title}{Blast wave in a nozzle for propulsive applications}.
\newblock \bibinfo{journal}{acta Astronautica} \bibinfo{volume}{3}, \bibinfo{pages}{141--156}.
\bibitem[{Wen et~al.(2023)Wen, Fan, Xu and Wang}]{wen2023numerical}
\bibinfo{author}{Wen, H.}, \bibinfo{author}{Fan, W.}, \bibinfo{author}{Xu, S.}, \bibinfo{author}{Wang, B.}, \bibinfo{year}{2023}.
\newblock \bibinfo{title}{Numerical study on droplet evaporation and propagation stability in normal-temperature two-phase rotating detonation system}.
\newblock \bibinfo{journal}{Aerospace Science and Technology} \bibinfo{volume}{138}, \bibinfo{pages}{108324}.
\bibitem[{Wu and Lee(2015)}]{wu2015stability}
\bibinfo{author}{Wu, Y.}, \bibinfo{author}{Lee, J.H.}, \bibinfo{year}{2015}.
\newblock \bibinfo{title}{Stability of spinning detonation waves}.
\newblock \bibinfo{journal}{Combustion and Flame} \bibinfo{volume}{162}, \bibinfo{pages}{2660--2669}.
\bibitem[{Xu et~al.(2024)Xu, Jin, Fan, Wen and Wang}]{xu2024numerical}
\bibinfo{author}{Xu, S.}, \bibinfo{author}{Jin, X.}, \bibinfo{author}{Fan, W.}, \bibinfo{author}{Wen, H.}, \bibinfo{author}{Wang, B.}, \bibinfo{year}{2024}.
\newblock \bibinfo{title}{Numerical investigation on the interaction characteristics between the gaseous detonation wave and the water droplet}.
\newblock \bibinfo{journal}{Combustion and Flame} \bibinfo{volume}{269}, \bibinfo{pages}{113713}.
\bibitem[{Yan et~al.(2021)Yan, Teng and Ng}]{yan2021effects}
\bibinfo{author}{Yan, C.}, \bibinfo{author}{Teng, H.}, \bibinfo{author}{Ng, H.D.}, \bibinfo{year}{2021}.
\newblock \bibinfo{title}{Effects of slot injection on detonation wavelet characteristics in a rotating detonation engine}.
\newblock \bibinfo{journal}{Acta Astronautica} \bibinfo{volume}{182}, \bibinfo{pages}{274--285}.
\bibitem[{Yoshizawa(1986)}]{yoshizawa1986statistical}
\bibinfo{author}{Yoshizawa, A.}, \bibinfo{year}{1986}.
\newblock \bibinfo{title}{Statistical theory for compressible turbulent shear flows, with the application to subgrid modeling}.
\newblock \bibinfo{journal}{The Physics of fluids} \bibinfo{volume}{29}, \bibinfo{pages}{2152--2164}.

\end{thebibliography}


\end{document}